\documentclass[12pt,preprint]{aastex}

\shorttitle{Dynamical Simulation of HD 141569A}
\shortauthors{Ardila et al.}

\begin{document}

\title{A Dynamical Simulation of the Debris Disk Around HD 141569A}


\author{
D.~R. Ardila\altaffilmark{1},
S.~H. Lubow\altaffilmark{2},
D.~A. Golimowski\altaffilmark{1},
J.~E. Krist\altaffilmark{2},
M. Clampin\altaffilmark{7},
H.~C. Ford\altaffilmark{1},
G.~F. Hartig\altaffilmark{2},
G.~D. Illingworth\altaffilmark{3},
F. Bartko\altaffilmark{4}, 
N. Ben\'{\i}tez\altaffilmark{1},
J.~P. Blakeslee\altaffilmark{1},
R.~J. Bouwens\altaffilmark{1},
L.~D. Bradley\altaffilmark{1},
T.~J. Broadhurst\altaffilmark{5},
R.~A. Brown\altaffilmark{2},
C.~J. Burrows\altaffilmark{2},
E.~S. Cheng\altaffilmark{6},
N.~J.~G. Cross\altaffilmark{1},
P.~D. Feldman\altaffilmark{1},
M. Franx\altaffilmark{8},
T. Goto\altaffilmark{1},
C. Gronwall\altaffilmark{9},
B. Holden\altaffilmark{3},
N. Homeier\altaffilmark{1},
L. Infante\altaffilmark{10}
R.~A. Kimble\altaffilmark{7},
M.~P. Lesser\altaffilmark{11},
A.~R. Martel\altaffilmark{1},
F. Menanteau\altaffilmark{1},
G.~R. Meurer\altaffilmark{1},
G.~K. Miley\altaffilmark{8},
M. Postman\altaffilmark{2},
M. Sirianni\altaffilmark{2}, 
W.~B. Sparks\altaffilmark{2}, 
H.~D. Tran\altaffilmark{13}, 
Z.~I. Tsvetanov\altaffilmark{1},   
R.~L. White\altaffilmark{2},
W. Zheng\altaffilmark{1}
\& A.~W. Zirm\altaffilmark{8}}

\altaffiltext{1}{Department of Physics and Astronomy, Johns Hopkins
University, 3400 North Charles Street, Baltimore, MD 21218.}

\altaffiltext{2}{STScI, 3700 San Martin Drive, Baltimore, MD 21218.}

\altaffiltext{3}{UCO/Lick Observatory, University of California, Santa
Cruz, CA 95064.}
\altaffiltext{4}{Bartko Science \& Technology, 14520 Akron Street, 
Brighton, CO 80602.}	
\altaffiltext{5}{Racah Institute of Physics, The Hebrew University,
Jerusalem, Israel 91904.}
\altaffiltext{6}{Conceptual Analytics, LLC, 8209 Woburn Abbey Road, Glenn Dale, MD 20769}
\altaffiltext{7}{NASA Goddard Space Flight Center, Code 681, Greenbelt, MD 20771.}
\altaffiltext{8}{Leiden Observatory, Postbus 9513, 2300 RA Leiden,
Netherlands.}
\altaffiltext{9}{Department of Astronomy and Astrophysics, The
Pennsylvania State University, 525 Davey Lab, University Park, PA
16802.}
\altaffiltext{10}{Departmento de Astronom\'{\i}a y Astrof\'{\i}sica,
Pontificia Universidad Cat\'{\o}lica de Chile, Casilla 306, Santiago
22, Chile.}
\altaffiltext{11}{Steward Observatory, University of Arizona, Tucson,
AZ 85721.}
\altaffiltext{12}{European Southern Observatory,
Karl-Schwarzschild-Strasse 2, D-85748 Garching, Germany.}
\altaffiltext{13}{W. M. Keck Observatory, 65-1120 Mamalahoa Hwy., 
Kamuela, HI 96743}

\begin{abstract}
We study the dynamical origin of the structures observed in the scattered-light images of the resolved debris disk around HD~141569A.  The disk has two conspicuous spiral rings and two large-scale spiral arms.  We explore the roles of radiation pressure from the 
central star, gas drag from the gas disk, and the tidal forces from two nearby stars in creating and maintaining these structures.  The 
disk's color, scattering function, and infrared emission suggest that submicron-sized grains dominate the dust population observed in 
scattered light. CO observations indicate the presence of up to $60~M_\earth$ of gas. The dust grains are subject to the competing effects 
of expulsive radiation pressure ($\beta > 1$, where $\beta$ is the ratio of the radiation and gravitational forces) and retentive gas 
drag.  We use a simple one-dimensional axisymmetric model to show that the presence of the gas helps confine the dust and that a broad ring of dust is produced if a central hole exists in the 
disk.  This model also suggests that the disk is in a transient, excited dynamical state, as the observed dust creation rate applied over the age of the star is inconsistent with submillimeter mass measurements. We model in two dimensions the effects of a fly-by encounter between the disk and a binary star in a prograde, parabolic, coplanar 
orbit.  We track the spatial distribution of the disk's gas, planetesimals, and dust.  We conclude that the surface density distribution reflects the planetesimal distribution for a wide range of parameters. Our most viable model features a disk of initial radius 
400~AU, a gas mass of $50 M_\earth$, and $\beta = 4$ and suggests that the system is being observed within 4000~yr of the fly-by
periastron.  The model reproduces some features of HD~141569A's disk, such as a broad single ring and large spiral arms, but it does not 
reproduce the observed multiple spiral rings or disk asymmetries nor the observed clearing in the inner disk. For the latter, we consider the effect of a $5~M_{Jup}$ planet in an eccentric orbit on the planetesimal distribution of HD~141569A.

\end{abstract}


\section{Introduction}
Debris disks around main-sequence stars are dusty, optically thin, and gas-poor.  Radiation pressure (RP) and Poynting-Robertson (PR) drag 
eliminate dust grains on timescales shorter than the stellar age, so the observed dust must be continuously replenished by collisions 
among, or evaporation of, planetesimals \citep{bac93}.  The {\it Infrared Astronomical Satellite (IRAS)} and the {\it Infrared Space 
Observatory (ISO)} revealed over 100 stars with far-infrared excesses indicating the presence of debris disks.  However, spatially resolved images 
of these disks are relatively rare:  only about a dozen debris disks have been resolved since the early 1980s (see \citealp{zuc01} and references therein). 

The resolved disks are not featureless.  They frequently display warps, spiral structures, and other azimuthal and radial asymmetries.  
Different mechanisms have been suggested to explain these features.  They may be caused by interactions between the dust and the gas in the 
disk (\citealp{tak01}, hereafter TA01), the formation of small planets (\citealp{ken04}), the dynamical forces of embedded planets (\citealp{oze00}), or 
stellar-mass companions (\citealp{lar01}; \citealp{aug04}; \citealp{qui05}).  Whatever their cause(s), the features observed in the dust 
disks provide insight into the characteristics of the unseen planetesimal population and illuminate the dynamical processes of young 
planetary systems. 

HD~141569A (A0~V; age 5$\pm$3 Myr, \citealp{wei00}) has a resolved circumstellar debris disk.  Its {\it Hipparcos} distance is 99~pc from the Sun and has co-moving M2~V and M4~V companions separated by 1.4'' and located at distances of 7\farcs55 and 8\farcs93, respectively 
(\citealp{wei00}; \citealp{aug99}).  The disk was first resolved in scattered light by \citet{wei99} and \citet{aug99} using the {\it Hubble 
Space Telescope's (HST's)} Near-Infrared Camera Multi-Object Spectrometer (NICMOS).  \citet{mou01} imaged the disk with {\it HST's} Space 
Telescope Imaging Spectrograph (STIS) and noticed strong brightness asymmetries.  These observations have recently been complemented by 
images from {\it HST's} Advanced Camera for Surveys (ACS) (\citealp{cla03}, hereafter C03).  Together, these scattered-light images reveal 
very complex structure, including an inner clearing within 175~AU of the star, a bright spiral ``ring'' with a sharp inner edge from 175 to 
215~AU, a faint zone from 215 to 300~AU, and a broad spiral ``ring'' from 300 to 400~AU (C03).  These distances are measured along the 
projected disk's southern semimajor axis.  (For convenience, we will hereafter refer to the two tightly-wound spiral rings simply as ``rings.'')  
C03 also observed two low-intensity, large-scale spiral arms in the outermost part of the disk, that they attributed to tidal interaction with the M dwarf companions.

The influence of the M dwarf companions on the disk's morphology was first discussed by \citet{wei00}. They argued that if all three stars 
were coplanar with the disk, the system would not be stable.  Given its young age, however, the system may be bound.  With this assumption,
\citet{wei00} concluded that resonant interactions between the companions and the disk do not account for the disk structure.  Assuming that
the companions revolve around the primary star in a highly eccentric orbit, \citet{aug04} constructed dynamical models of the optical disk 
comprising dust grains that respond to the stellar gravitational field.  They assumed that the grains are large enough that RP, PR drag, and gas drag
are unimportant.  They also ignored the presence of the large-scale spiral arms reported by C03.  Their
model reproduces the general appearance of the brightness asymmetries observed in the disk.  \citet{qui05} constructed hydrodynamic models of HD~141569A's gas disk without dust, and concluded also that the companions lie in an eccentric orbit.  Their 
models reproduced the large-scale spiral arms (even after several periastron passages) and some other disk asymmetries. 

TA01 modeled the interaction between disks of gas and intermediately-sized ($\gtrsim 8~\mu$m) dust grains, and \citealt*{lec98} 
considered disks with very small particles.  It is not clear that the conditions explored by any of these models is the correct one for the scattered light images of the disk around 
HD~141569A.  In \S\ref{obscons}, we argue that submicron-sized grains account for most of the scattering opacity at optical wavelengths.
Such small grains are subject to RP which quickly expels them from the disk.  The gas detected in the disk \citep{zuc95} tempers somewhat 
the short ``blow out'' timescales, but its mass is not enough to dominate the dynamics of the dust (\S\ref{conf}).  In this paper, we explore 
the effects of strong RP, gas drag, and gravity on small dust grains, using HD~141569A's disk as a benchmark.  We simultaneously track the 
behavior of three different populations -- the unseen planetesimals controlled by gravity, the gas controlled by gravity and volumetric 
fluid forces, and the dust controlled by gravity, RP, and gas drag.  The insights obtained from this analysis should be applicable to other
systems where RP is important.

In \S\ref{obscons}, we review the observational constraints that every model of the system should satisfy.  We consider evidence that indicates
the dominance of submicron-sized grains and we set limits on the amount of gas in the disk.  In \S\ref{conf}, we explore the interaction
between gas and dust by means of a one-dimensional model.  We show that the gas slows the outward motion of the dust and can produce broad 
ring-like structures. In \S\ref{model}, we present two-dimensional dynamical simulations of the interactions between planetesimals, dust 
grains, gas, and the three stars.  Assuming that the two M dwarf companions are co-moving but unbound, we find that a recent parabolic 
fly-by causes some of the observed disk structure, including the large-scale arms.  However, our models produce a more disorganized disk 
than is observed.  We argue that the central hole in the disk can be produced by a planet with a mass a few times that of Jupiter (or,
alternatively, some number of smaller planets) in a highly eccentric orbit about the primary star.

\section{Observational Constraints \label{obscons}}

\subsection{Grain Size}
\begin{figure}
\plotone{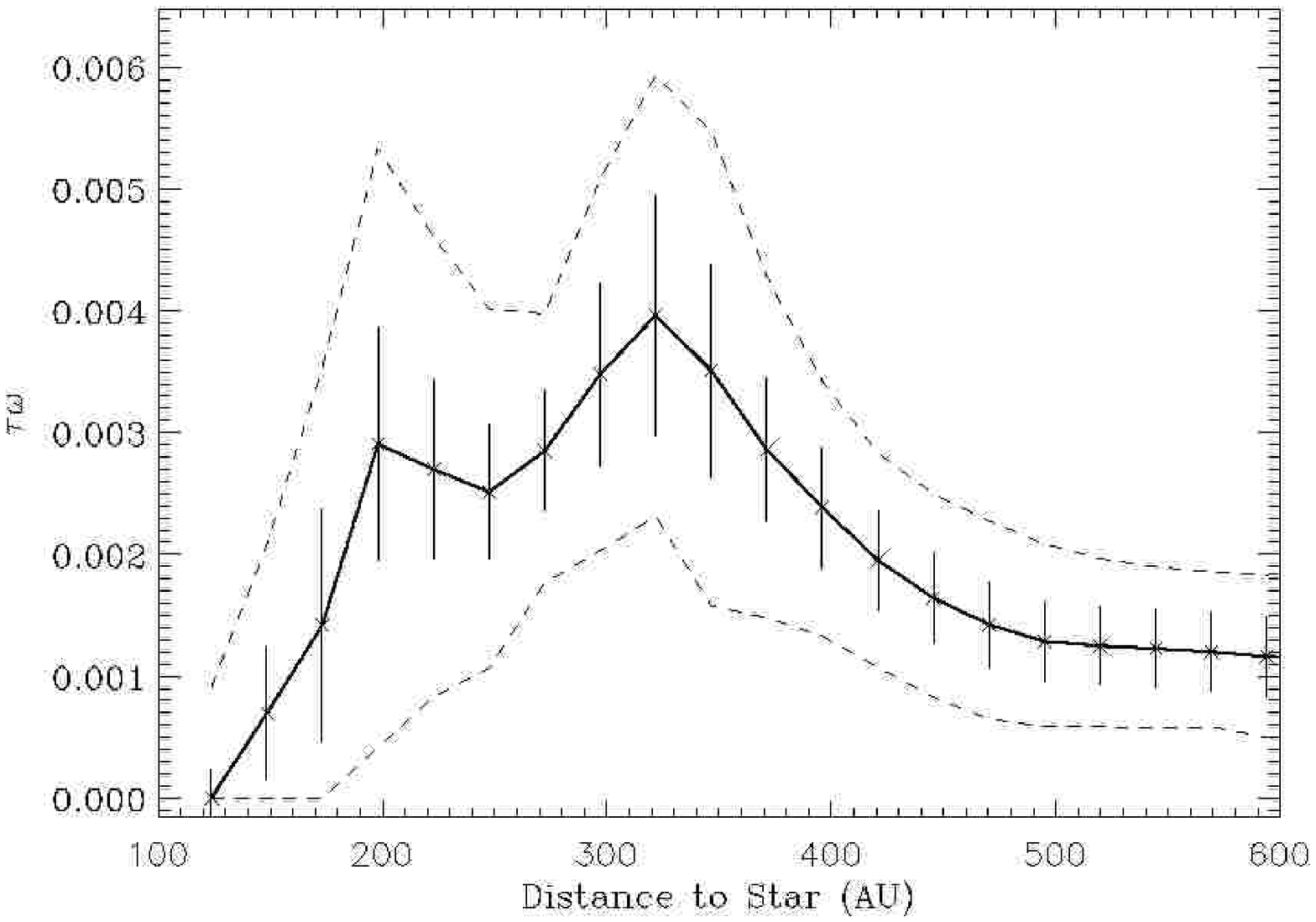}
\caption{\label{profiles}Optical depth profile (proportional to the surface density) of the disk shown in Figure~5 of C03, derived from the median values of concentric 
annuli centered on the estimated geometric center of the spiral rings.  The error bars indicate the standard deviation and the dashed lines indicate the upper and lower values in each annulus.}
\end{figure}

Figure \ref{profiles} shows the optical depth profile of the disk, taken from Figure 5 of C03.  This profile represents
the median values of concentric annuli centered on the primary star.  Drawn from a composite of F435W (ACS $B$ band) and F606W (ACS 
broad-$V$ band) images, it is roughly indicative of the optical characteristics of the disk at $0.5~\mu$m. Power-law fits to the inner and outer edges of the profile (using as errors the standard deviation at every radius) give $r^{5\pm3}$ for $r<200$~AU and $r^{-2.8\pm0.6}$ for $320<r<500$~AU. Beyond 500~AU, the average azimuthal brightness seems to be dominated by the light from HD~141569BC. The optical depth profile indicates that the amount of mass in $\sim0.5~\mu$m-sized grains is $\sim 0.01$M$_\earth$ (assuming a constant dust opacity of $2\ 10^4$ cm$^2$/gr and albedo of $\sim 0.5$. See \citealt{woo01}). 

The disk is occulted by ACS's coronagraphic mask within $\sim150$~AU.  \citet{fis00} and \citet{mar02} detected mid-infrared thermal 
emission within this radius and \citet{mar02} concluded that the optical depth at $1.1~\mu$m decreases by a factor of $\sim4$ within this region.  
\citet[hereafter LL03]{li03} modeled the thermal spectrum of the disk by assuming a density profile 
similar to the one shown in Figure \ref{profiles}.  Their model suggests that the reduced emission within $\sim$150~AU is due to
reduced dust density and not a change in the scattering properties of the grains. 
 
C03 reported that the disk is redder than the star, with color excesses of $\Delta(B-V)=0.21$ and $\Delta(V-I)=0.25$.  They also reported no 
color variation as a function of distance from the star (which supports our use of a constant opacity in the mass calculation above).  C03 inferred that the disk's colors are consistent with astronomical-silicate grains 
having a size distribution of $s^{-3.5}$ and a minimum radius of $s \sim 0.4~\mu$m.  \citet{aug04} used the same color information to derive 
lower size limits between $\sim0.1~\mu$m and $3.1~\mu$m.  The existence of such small grains is also implied by the small scattering asymmetry 
factor ($0.15 < g < 0.25$\footnote{Because of a typo, the value of $g$ for the HD~141569A disk is quoted as $g=0.25-0.35$ in C03}) of the Henyey-Greenstein function derived by C03.  The multicomponent models by LL03 
suggest that the minimum grain size is between $0.1~\mu$m and $10~\mu$m: the smaller limit produces too little emission in the 
{\it IRAS} 60 $\mu$m band and the larger limit produces too much emission in the {\it IRAS} 60 $\mu$m and 100 $\mu$m bands.  Taken together, these arguments imply that there is a population of grains whose radii
may extend down to $0.1~\mu$m.

\begin{figure}
\plotone{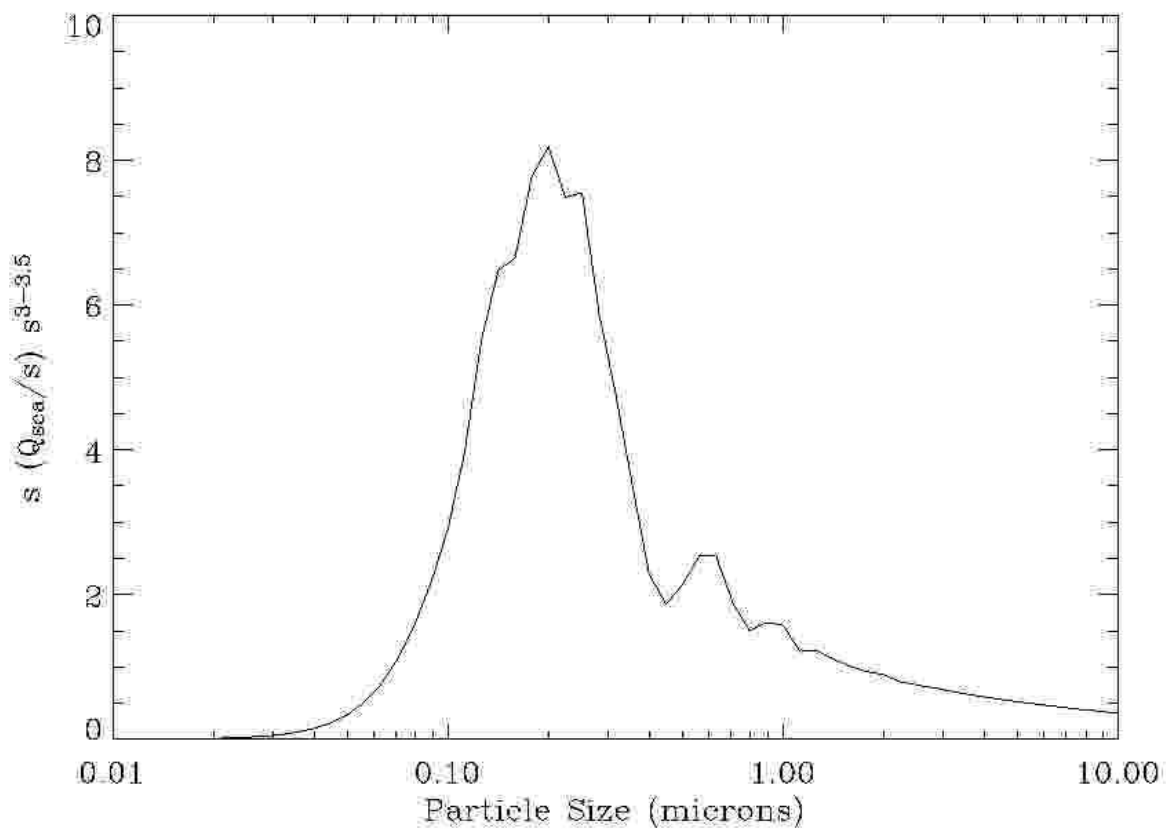}
\caption{\label{draine}Scattering opacity at $0.5~\mu$m as a function of grain size.  The scattering opacity is proportional to the 
product the mass opacity (the scattering cross section, $\sigma_{sca}=Q_{sca} \pi s^2$, divided by the particle mass, $\propto s^3$) 
weighted by the number of particles ($\propto s^{-3.5}$) and the particle mass ($\propto s^3$), where $s$ is the particle size. The 
scattering efficiency $Q_{sca}$ is that of a compact astronomical silicate \citep{lao93, dra84}. To conserve the area under the 
curve with a logarithmic abscissa, the ordinate has been multiplied by $s$.}
\end{figure}

The combined influences of the reduced scattering efficiency ($Q_{sca}$) of small grains and the small number of large grains (assuming a size distribution going as  $s^{-3.5}$) produce sharply peaked scattering opacity, which indicates that the grain sizes responsible for the optical images of the disk are of the order of the wavelength of observation.  Figure \ref{draine} shows the scattering opacity at 
$0.5~\mu$m as a function of grain radius for astronomical silicate grains \citep{lao93, dra84} larger than $0.01~\mu$m.  The scattering opacity 
is expressed as the scattering cross section per unit mass ($\propto \sigma_{sca}s^{-3}$, where $\sigma_{sca} = Q_{sca} \pi s^2$), weighted 
by mass and number of grains at each radius $s$ \citep{miy93}.  The function is centered at $0.2~\mu$m with a characteristic width of $\Delta s 
\sim 0.3~\mu$m.  The sharp decrease in the scattering opacity of grains larger than $0.2~\mu$m is determined by the behavior of the first 
scattering peak in $Q_{sca}$.  In the limit of large grains, the scattering opacity decreases more slowly ($\sim s^{-0.5}$, for intervals of constant $ds/s$).  For this dust model, $1~\mu$m grains scatter $\sim 8$ times less than $0.2~\mu$m grains, and $\sim 3$ times more than $10~\mu$m 
grains.  The location and width of the peak in the scattering opacity depends on the surface properties of the grains and the exact mixture 
of silicates, ices, and empty space.  For example, increasing porosity moves the peak to smaller sizes and less-reflecting grains broaden 
the peak and move it to larger sizes \citep{boh83}. 

If -- as suggested by the disk colors, the phase of the scattering function and the spectral energy distribution -- submicron-sized grains 
exist in the disk, their dynamics are strongly influenced by RP.  According to TA01, the ratio ($\beta$) of the forces from RP and 
gravity is $\sim4$ for a $1~\mu$m dust grain, and $\beta \sim 20$ for $s\sim0.05~\mu$m.  For the very porous grains assumed by LL03, 
$\beta\sim12$ for $s\sim1~\mu$m.  Formally, and in the absence of gas, grains with $\beta>0.5$ (``$\beta$-meteoroids'') are expelled from 
the disk.  Grains with large $\beta$ will acquire a radial velocity on the order of their Keplerian velocity in $T_{esc} \sim (P/2\pi ) 
/ \beta$, where $P$ is the orbital period at a given position (TA01).  At the edge of a 500~AU disk around a 2.3 M$_\sun$ star,  
$T_{esc} \sim 300$ yr for $\beta=4$.  Such short expulsion timescales are somewhat mitigated by the effects of another force: the gas drag.


\subsection{Gas Content}

\citet{zuc95} measured the gas content of HD~141569A's disk using the detected $J=2\rightarrow1$ transition for $^{12}$CO and the upper limit 
on the same line for $^{13}$CO.  They assumed that the star's distance was 200~pc, its disk had a radius of 130~AU, and that H2/CO $\sim$ 10000.
They calculated gas masses between 20 and $460~M_\earth$ for optically thin and optically thick limits, respectively.  Assuming that 
the gas and dust disks are coincident, we recompute the gas content using the observed extent of the optical disk and the measured 
{\it Hipparcos} distance to HD~141569A of 99 pc.

\begin{figure}
\plotone{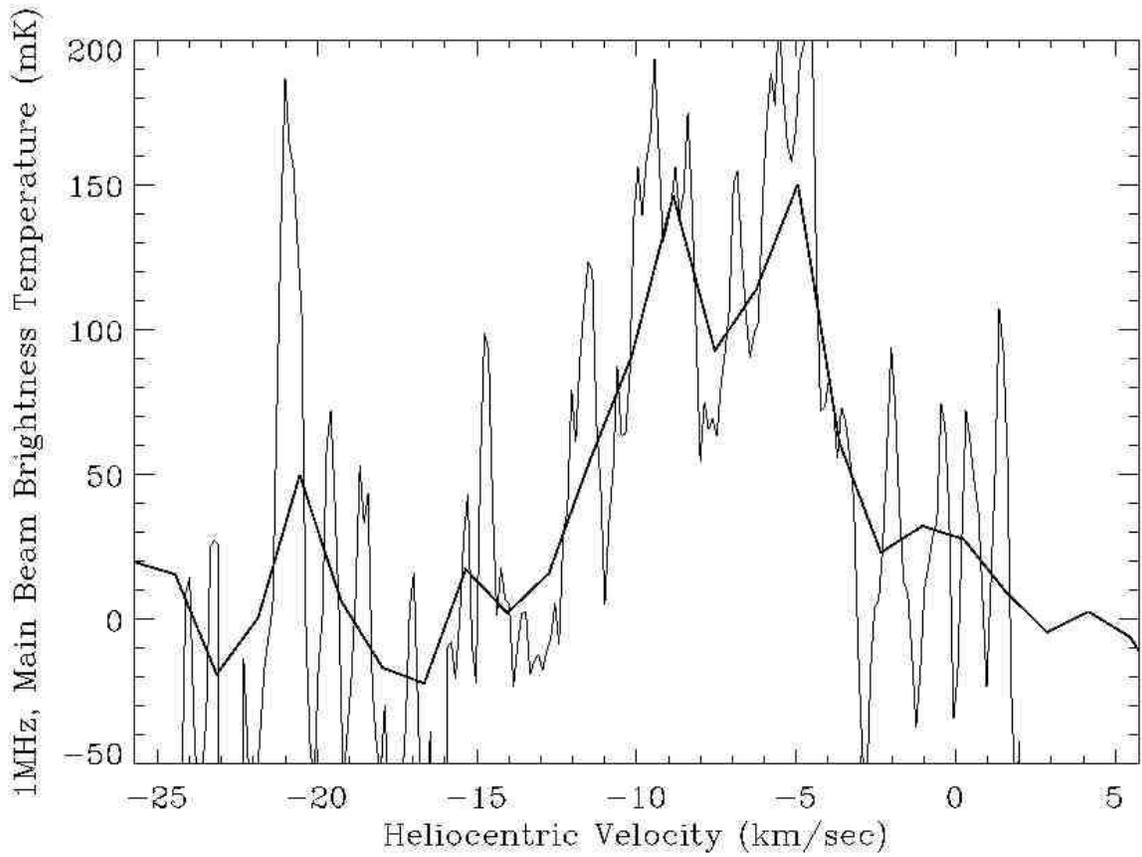}
\caption{\label{zuckerman}Spectra of $^{12}$CO $J=2\rightarrow1$ transition with resolutions of 1~MHz {\it (thick curve)} and 100~KHz 
{\it (thin curve, after three-point smoothing)} from \citet{zuc95}.  Although the depth of the central depression is $\sim 1\sigma$, it
is present in both spectra.}
\end{figure}

\citet{zuc95} have generously provided us with their unpublished digitized CO spectra, which are reproduced in Figure \ref{zuckerman}.  The spectra show a 
double peak characteristic of sharp disks, although the depths of the central line-emission in both measurements are a little over $1\sigma$ 
per resolution element.  Assuming an disk inclination of $55^{\circ}$ (C03), the velocity difference between the peaks is 
$3.8\pm1.0$~km~s$^{-1}$, which implies emission at $380_{-180}^{+120}$~AU.  (The upper limit is dictated by the size of the optical disk.) 
Assuming H2/CO $\sim$ 4000 (\citealp{lac94}), the optical depth of the central line is $\tau=1.5\times 10^6 \ (\Sigma/gr \ cm^{-2}) \ (5.5K/T)^{5/2}$, 
where $\Sigma$ is the surface density and $T$ is the excitation temperature of the gas \citep{bec93}.  Nondetection of the $^{13}$CO line 
yields an upper limit of $T = 30$~mK, so the opacity of the $^{12}$CO line is uncertain.  If the $^{12}$CO line is optically thick, then we 
obtain a gas mass of $\lesssim 60~M_\earth$.  If the line is optically thin and fills a disk of radius 380~AU, we obtain a gas mass of 
$0.02~M_\earth$.  In each case, the gas is assumed to be in local thermal equilibrium (LTE). 



\section{The Dynamics of High $\beta$ Particles\label{conf}}

We now explore the interaction between small dust grains, gas, and the radiation and gravitational fields of HD~141569A, while ignoring the presence of the binary companions.  Most of 
the results of this axisymmetric treatment should be applicable to other debris and gas disks in which RP is important. The regime that we explore here is that of modest amounts of gas (up to 50 M$_\earth$) and not very large values of $\beta$ ($\beta\sim4-10$). Our calculations are complementary to those of \citet{lec98}, who consider $\beta\gtrsim 100$ and TA01, who consider $\beta\lesssim1$. Notice that \citet{lec96} has shown that if the observed dust particles all have $\beta < 0.5$, a single ring of planetesimals in a gasless disk will produce a dust disk with surface density given by $r^{-3}$, as observed here.

We solve the equations of motion for a dust grain, assuming that it is subject to RP, PR and gas drag. The gas is assumed to be in a circular orbit and we assume the same dust model as TA01. The force of the gas on the dust is given by 

\begin{equation}
\label{fg}
{\bf F_g}=-\pi \rho_g s^2 \left(v_T^2+\Delta v^2\right)^{1/2}{\bf \Delta v},
\end{equation}

\noindent
where $\rho_g$ is the gas density, $s$ is a dust grain radius, $v_T$ is 4/3 times the thermal velocity of the gas, and ${\bf \Delta v}$ is the relative 
velocity of the dust with respect to the gas. To determine the trajectory of a dust grain, we assume that the surface density of the gas (as well as that of planetesimals) 
decreases radially as $r^{-1.5}$.  This function is characteristic of the dust surface density of optically thick protoplanetary disks  
\citep{ost95}.  In other words,we assume that the gas and planetesimal density profiles of the optically thin disks are the same as the dust density profile of the optically thick disks. We further assume (unless otherwise stated) that the gas is in LTE, with temperature 

\begin{equation}
T_g=278 \left(\frac{L_*}{L_{\sun}}\right)^{1/4}{r_{AU}}^{-1/2}~K \end{equation}
 
\noindent 
where $L_* = 22.4 L_{\sun}$ is the stellar luminosity \citep{bac93}. This implies that $H$, the vertical scale of the gas disk, is given by
\begin{equation}
\label{lte}
\frac{H}{r}=0.185 \left(\frac{L_*}{L_{\sun}}\right)^{1/8}\left(\frac{r}{1000~AU}\right)^{1/4}.
\end{equation}

\begin{figure}
\plotone{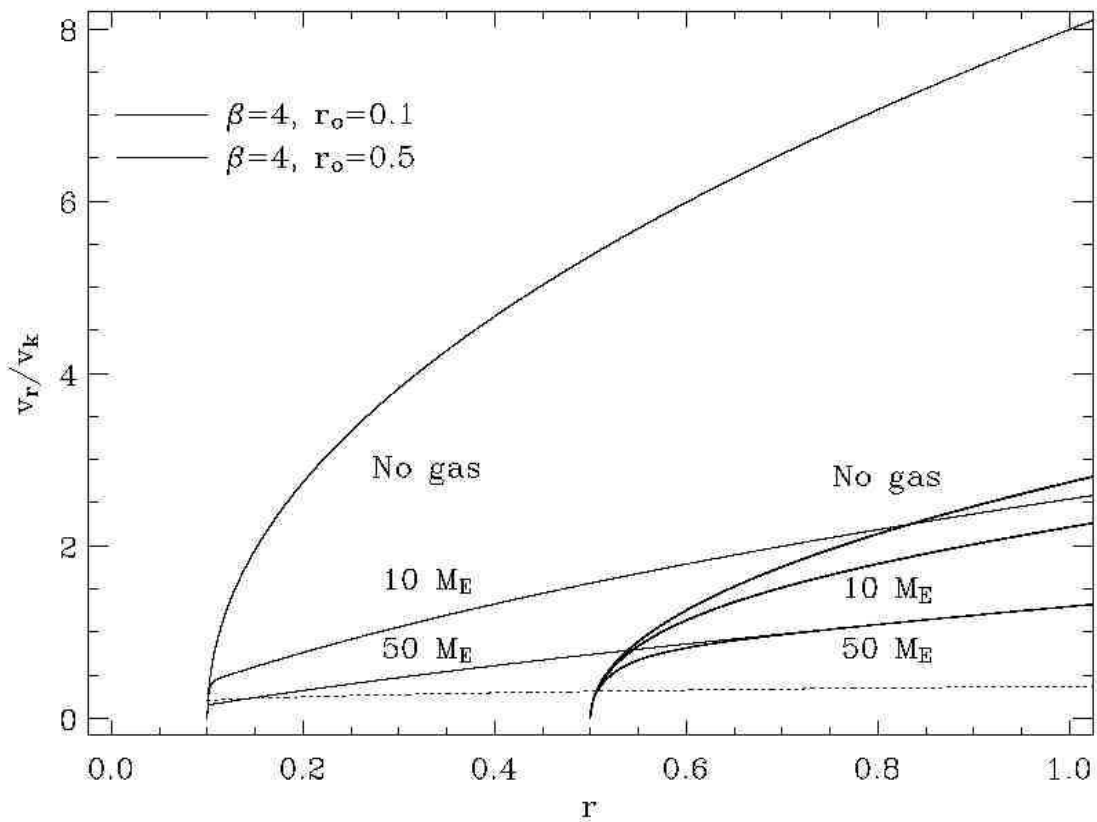}
\caption{\label{rad_vel}Radial velocity (divided by the local Keplerian velocity) as a function of radii, for $\beta=4$ dust particles launched from two different points: $r_o=0.1$ (thin lines) and $r_o=0.5$ (thick lines). The behavior of the dust particles with different amounts of gas is shown: no gas (highest curve), 10 M$_\earth$, and 50 M$_\earth$ (lowest). The dotted line is the thermal sound speed of the gas. }
\end{figure}

Without gas, the radial velocity of a dust grain is given by

\begin{equation}
v_r(r_o,r)=\sqrt{2}v_o\left[\beta-\frac{1}{2}-\frac{r_o^2}{2r^2}+\frac{r_o}{r}(1-\beta)\right]^{1/2},
\end{equation}

\noindent
where $r_o$ is the radius at which the dust grain is created and $v_o$ is the Keplerian velocity at $r_o$. In general, there is not an analytic expression for the radial velocity of a dust grain in the presence of gas. For the range of parameters we are considering here, the dust grain speeds are supersonic and almost radial at the outer edge of the disk (Figure \ref{rad_vel}). The dimensionless stopping time (the stopping time in units of the inverse local Keplerian frequency) at the edge of the disk is given by (TA01):

\begin{equation}
T_s\sim \frac{4 \rho_d s v_K}{3 \rho_g r v_r} \sim 86.6(\frac{1}{\beta})(\frac{M_\earth}{M_{gas}})(\frac{1}{v_r/v_k})
\end{equation}
 
\noindent
which for the parameters considered here is $\lesssim1$. Notice that, unlike in the case considered by \citet{lec98}, the radial velocity shown here does not have a deceleration region, because in our case the gas surface density decreases with radius: the radiation pressure always dominates the dynamics of the dust.

\begin{figure}
\plotone{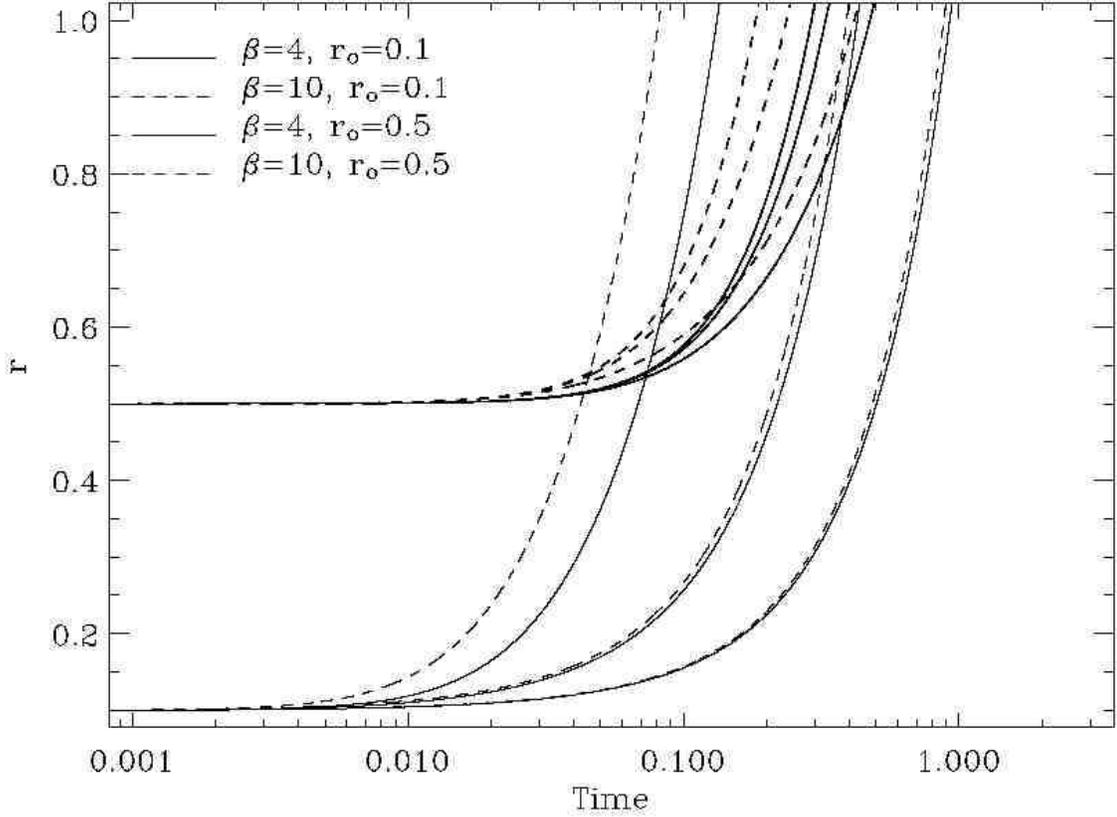}
\caption{\label{time_it_takes}Position as a function of time for dust particles released at two different positions: 0.1 (thin lines) and 0.5 units (thick lines), for a unit disk of gas in thermal equilibrium around a 2.3 M$_\sun$ star. The unit of time is such that the orbital period at the outer edge of the disk is 2$\pi$. (If the disk is 1000 AU in radius, one time unit is 3325 yrs.)  The solid lines trace particles with $\beta=4$, while the dashed lines trace particles with $\beta=10$. For each set, three lines are shown. From the left: no gas, 10 M$_\earth$ and 50 M$_\earth$. Notice that for large amounts of gas there is little difference between particles with $\beta=10$ (the dashed lines) and $\beta=4$ (the solid lines).}
\end{figure}

The effect of gas on the grains depends on where the grains were created. Figure \ref{time_it_takes} shows that for a fixed value of $\beta$ and without gas, particles that start closer to the star soon overtake particles farther out. This changes with the presence of gas, because the drag force decreases outward, and for $\gtrsim 50 $M$_\earth$ masses of gas, the particles created closer in will not overtake those created farther out. Figure~\ref{time_it_takes} also shows that the gas mass, not the grain size, determines the escape timescale: there is not much difference between $\beta=4$ and $\beta=10$ for $50~M_\earth$ of gas, because both the gas drag and the RP are proportional to $\beta$.

If one assumes a rate of dust generation, the results of these calculations can be used to predict the steady-state dust surface density in an isolated disk. In such a state, grains are continuously generated by collisions and then blown away by RP.  The dust surface density (assuming conservation in the number of particles) is given by 

\begin{equation}
\label{sigma}
\Sigma(r) \propto \int_{r_i}^{r} \frac{dN(r_o)}{v_r(r_o,r)r},
\end{equation}

\noindent
where $r_i$ is the inner radius of dust generation, $r_o$ is the position at which the grains are created, $dN(r_o)$ is the number of grains 
created at $r_o$, and $v_r(r_o,r)$ is the radial velocity at $r$ of grains created at $r_o$.  The dust creation rate per unit volume is taken to be v$_{rel} \times \sigma_{coll} \times \rm{N_p}^2$, where $\rm{N_p}$ is the number of planetesimals per unit volume, $\sigma_{coll}$ is the collision cross section and v$_{rel}$ is the relative velocity between the planetesimals, which is proportional to the dimensionless planetesimal disk thickness ($H_p/r$) times the Keplerian velocity \citep{the03}. This implies 
\begin{equation}
 dN(r_o) = v_{rel}\ \sigma_{coll} \ \rm{N_p}^2 \ \rm{H_p} \ 2\pi r_o dr_o \propto \Omega_k \ \Sigma_p^2  \ 2\pi r_o dr_o
\end{equation}
\noindent
where $\Sigma_p$ is the surface density of planetesimals at the creation point, which we assume to be proportional to $r^{-1.5}$.

In practice, this simple prescription (the ``quadratic'' dust generation prescription) may be affected by uncertainties in the planetesimal size distribution and in the relative velocities among the the planetesimals. To explore the sensitivity of the results to the exact generation mechanism we also consider a ``linear'' dust generation prescription in which $dN(r_o) \propto \Sigma_p r_o dr_o$. 

\begin{figure}
\plotone{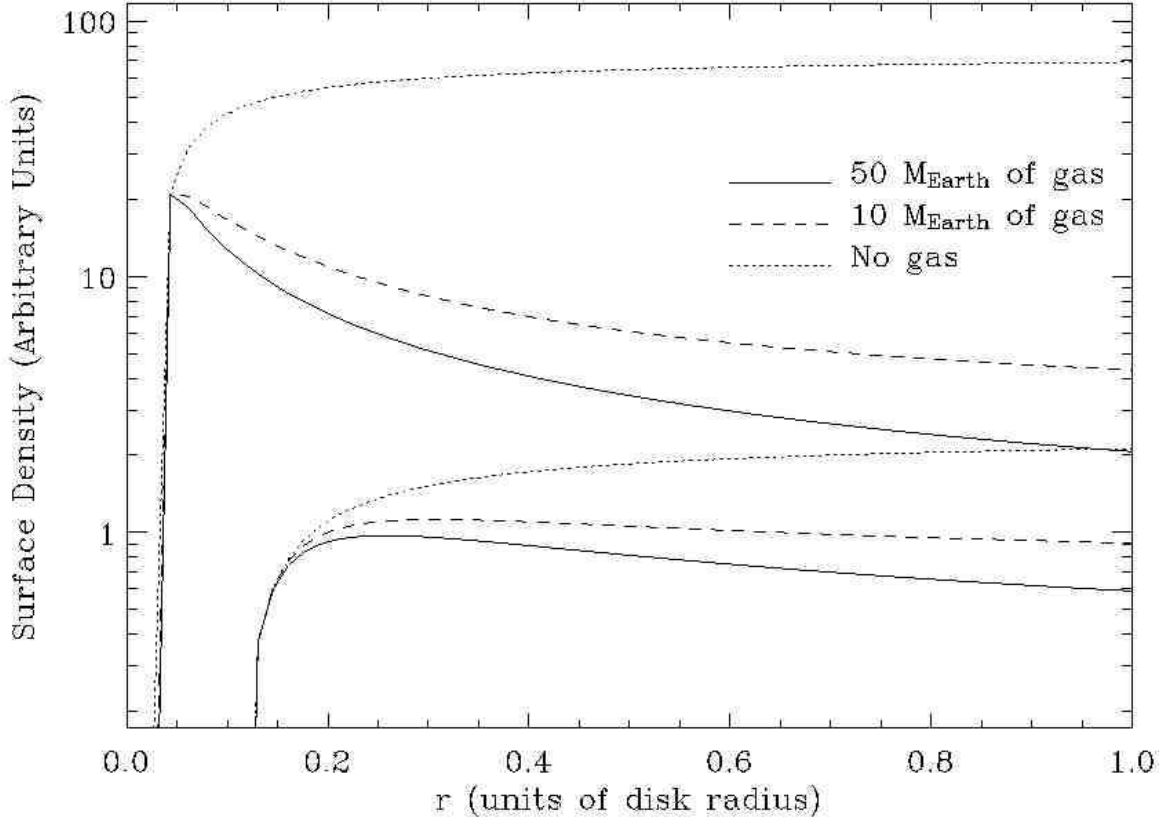}
\caption{\label{1d_model}Steady-state dust surface density as a function of distance for an isolated disk, with dust particles ($\beta=4$, $s \sim 1 ~\mu$m according to TA01 or $s\sim 3 ~\mu$m for LL03) for the linear model (in which the rate of generation is proportional to the planetesimal surface density). Two sets of surface densities are shown: one for a disk in which the innermost dust generation radius is 0.1 and another for which it is 0.01. The dotted lines indicate the surface density with no gas; the dashed lines show the surface density with 10 M$_\earth$ of gas; the solid lines show the surface density with 50 M$_\earth$ of gas. The gas is in LTE.}
\end{figure}

For these calculations we assume that the planetesimals and the gas are well mixed and therefore $H_p=H$. The resulting steady-state surface densities are shown in Figures~\ref{1d_model} and \ref{1d_model2}. In the figures, the unit length is 1000 AU. Figure \ref{1d_model} shows that even modest amounts of gas have a significant effect on the surface-density profiles.  The amounts of gas considered (10 and $50~M_\earth$) are not enough to confine the dust grains created at distances larger than 0.01, but gas drag sharpens the profiles, as it slows down the dust particles. Notice that if dust creation starts very close to the central star, the resultant surface-density profiles are effectively featureless. Figures~\ref{1d_model} and \ref{1d_model2} also show that a broad ring of dust is produced by truncating the dust creation within a certain radius.  Just outside this limit, the number and surface density of planetesimals, and hence dust grains, are large.  Far from the creation limit, the 
surface density of the dust decreases because of decreasing surface density of planetesimals and the geometric dilution of ``blown out'' dust 
grains created at smaller radii.  Thus, if a mechanism exists for preventing the creation of dust at small radii (for example, a planet that clears 
out the parent planetesimals), the result is, very naturally, a broad ring of dust at larger radii. 

The rate at which dust is lost from the disk depends on the rate of planetesimal erosion.  In principle, one can use use the former to estimate the latter. LL03 calculated the amount of mass lost from HD~141569A's disk, assuming no gas, a grain-size distribution $\propto s^{-3.3}$, $\beta$=1 for grains 
of all sizes, and a constant rate of mass loss throughout the age of the system.  They also assumed a surface-density profile for the dust that is slightly different than the one shown in Figure~\ref{profiles}.  They concluded that $39~M_\earth$ of solid material have been lost, in particles with $1<s<10 ~\mu$m, from the disk due to RP and PR drag.  

To examine the effect of the gas on the rate of mass loss, we consider a disk of radius 500~AU about HD~141569A and no binary companions.  
Figure~\ref{timescale} shows the time required for small dust grains to move from a given radius to the edge of the disk for different values of gas 
mass and temperature.  The presence of just $10~M_\earth$ of gas triples the escape time from 200~AU, and a colder disk confines the dust more than a 
hotter one. (Equation~\ref{fg} shows that, neglecting $v_T$, a colder -- denser -- gas disk, produces more drag). Assuming a grain-size distribution $\propto s^{-3.5}$, with $50~M_\earth$ of gas in LTE, varying escape times, the TA01 dust model and a constant rate of
mass-loss, we determine that $\sim700~M_\earth$ of solid material have been lost over the $\sim 5$~Myr age of the disk, in particles with sizes $0.1<s<8 ~\mu$m. Without gas, $\sim1200~M_\earth$ of solid material would have been lost. The difference with LL03 is due to the fact that in our calculation different size particles have different blow-out timescales.

However, neither LL03's estimate nor 
ours can be correct.  Because small grains ($\lesssim 10~\mu$m) are quickly expelled from the disk, they must be the dominant component of the
lost mass.  If the grain-size distribution is $\propto s^{-3.5}$ for grains with radii $< 1$~mm, then grains up to 10 $~\mu$m compose only ten percent of the disk mass. The bulk of the dust mass comprises large grains that are eliminated in much longer timescales.  If the collisional processes responsible for the 
replenishment of small grains also produce large grains, then submillimeter observations should detect $\sim10^4~M_\earth$ of dust, which greatly
exceeds the amount of $\sim2~M_\earth$ measured by \citealt*{syl01}. 

One possible solution to this discrepancy is that the dust is not generated with a power-law size distribution, but with a distribution weighted 
toward small particles.  The collisional-evolution models of \citealt*{the03} suggest that, although large variations in the dust distribution are possible,
most of the mass produced by the collisions is in large particles.  These models assume that the collisional timescale is smaller than any other timescale in the disk. 

Another possible solution is that the current rate of dust creation is larger than it has been in the past. This possibility is consistent with the idea that the disk has been recently stirred by a close encounter with a companion or unbound star. In section \ref{model} we show that a parabolic encounter would occur over timescales on the order of $10^3$ yrs. Over this time the disk would lose (according to the model above) $\sim1~M_\earth$ of solids in small particles, or  $\sim10~M_\earth$ in particles up to 1~mm in size, close to the measured value. Observationally, the presence of  $\sim0.01~M_\earth$ of solids in $0.5 ~\mu$m-sized particles (Section~\ref{obscons}) suggests that there is currently on the order of $0.1~M_\earth$ in particles up to 8 $\mu$m in size, with the exact number depending on the assumed value of the opacity and dust size distribution.

Furthermore, these estimates support the idea of a single exciting event, like an encounter with an unbound companion, as opposed to repeated exciting events. In the models by \citet{aug04}, which assume a bound companion, repeated encounters are required to explain the brightness asymmetries in the debris disk.

\section{A Dynamical Simulation\label{model}} 

The presence of large-scale spiral arms suggests that the dynamical influence of the binary companions, HD~141569B and C, is important.  If they are 
coplanar with the disk, their center of mass is $\sim1250$~AU away from the primary star and their separation is $\sim275$~AU.  \citet{aug04} assumed 
that the companions are bound to HD~141569A in a very eccentric orbit.  Their dynamical models show that close encounters between the companions and the 
disk produced well-developed spiral rings within the disk after only a few periastron passages.  However, the models do not maintain large-scale spiral
arms after a few orbits.  Thus, within their model, the interaction geometry needed to create the spiral rings in the disk is not consistent with the
presence of the large-scale spiral arms. The hydrodynamic model by \citet{qui05} does produce spiral arms after repeated close encounters, because the viscosity and gas pressure help generate a spiral at every periastron passage.  However, the amount of gas 
actually present in the disk (\S2.2) discourages the notion that the dynamics of the dust conforms to that of the gas.  Figure \ref{time_it_takes} shows 
that, for a reasonable range of gas mass, $\beta$-meteoroid grains responsible for the scattered-light disk are impeded by gas drag, but they are not bound by the gas.  So the assumption by \citet{qui05} that gas dynamics control the observed structures in the dust disk is not appropriate.

\subsection{Assumptions and Methodology\label{assump}} 

We assume that the encounter between the disk and the binary companions is a parabolic fly-by, a situation known to produce spiral arms in dusty disks 
\citep{lar01}.  Such an encounter is consistent with the proper motion of the system.  Within an error box of 0\farcs1, the position of the companions 
relative to HD~141569A has not changed over the 60~yr baseline of observations noted by \citet{wei00}.  The largest expected relative motion is obtained 
when the system is currently at periastron and the encounter occurs in the plane of the sky.  In this scenario, the relative motion of the companions 
over 60~yr would be 0\farcs012, which is well inside the error box of the proper motion measurements.

Dust grains are created by collisions of planetesimals.  We assume that the planetesimals themselves experience gravity, but not RP or gas drag.  We 
also ignore self-gravity and the dynamical effects of the collisions.  Our simulation is performed in two phases.  First, we obtain the distribution of 
planetesimals and gas as a function of time.  Then, for each planetesimal configuration, we generate a dust distribution and follow it as a function of 
time.  We repeat this sequence for all planetesimal and gas configurations.  At any given time, the dust in the disk consists of dust created at that 
time and dust remaining from earlier times.

The model is two-dimensional.  We assign fixed masses of 2.3, 0.5, and $0.25~M_\sun$ to HD~141569A, B, and C, respectively.  We assume that 
the encounter is prograde, as a retrograde encounter fails to produce large-scale spiral arms.  The phase of the encounter is fixed so that, 
at periastron, the three stars are aligned, with HD~141569B between A and C.  (The results are mostly insensitive to the choice of phase.) 
Most of our simulations start with $10^4$ planetesimals. The inner dust destruction limit is set at $r > 0.1$, where the distance $r$ is measured relative to the 
separation of HD~141569A and HD~141569BC's barycenter at periastron.  We assume that the initial surface density of planetesimals declines 
as $r^{-1.5}$.  The outer radius of the planetesimal disk is a free parameter.  Figure~\ref{panel} shows the evolution of the planetesimals in a disk of unit radius.  We define $T=0$ as the time of periastron. The time unit is such that the period at unit distance is 2$\pi \sqrt(\frac{m_1+m_2}{m_1})=7.2$ time units, where $m_1=2.3 M_\sun$ and $m_2=0.75 M_\sun$ (the later is the sum of the companion masses). Notice that the particular fly-by configuration shown in Figure~\ref{panel} shows that some planetesimals can be captured by the companions and therefore that some of the IR excess measured by {\it IRAS} may be associated with them.

The total mass of gas is a free parameter.  We consider gas disks with $r > 0.01$, an initial surface-density profile $\propto r^{-1.5}$ and masses of 
0, 10, 50, and 100 M$_\earth$, before truncation by the encounter.  We model the gas evolution using the Smoothed Particle Hydrodynamics (SPH)
formalism \citep{mon92}, which allows one to track the motion of gas pseudo-particles using Lagrangian equations. Viscosity is parametrized according to Equation~4.2 of \citet{mon92}.  SPH permits the 
exploration of parameter space without a large investment of computer time, but with limited resolution at small scales. 

We explore different limits for the gas sound speed, which is parametrized by the disk opening angle, $H/r=c_s/r\Omega_K$, where $H$ is the 
scale height of the gas, $c_s$ is the speed of sound, and $\Omega_K$ is the Keplerian angular frequency at 
radius $r$.  We consider gas with $H/r=0.05$, 0.1, 0.2, and in LTE (Equation~\ref{lte}). \citet{bra04} has shown that the empirical opening angle of the $\beta$~Pictoris gas disk is $H/r\sim0.28$, which is close to the expected 
LTE value at 1000~AU.  Figure \ref{panel_gas} shows the evolution of gas particles for the case of $H/r = 0.1$.

To simulate the dust grains, we consider the linear (particles created with $r>0.1$) and quadratic (particles created with $r>0.15$) dust-creation prescriptions described in \S\ref{conf}. Particles are created every 0.01 time units. The ratio $\beta$, which couples the dust and the radiation, is also a free parameter.
We explore two values: (1) $\beta=4$, which corresponds to $s\sim1~\mu$m (TA01) or $s\sim 4~\mu$m (LL03), and (2) $\beta=10$, which 
corresponds to $s\sim 0.3~\mu$m (TA01) or $s\sim1~\mu$m (LL03).  We assume a single value of $\beta$ per simulation, which is 
equivalent to assuming that the scattering opacity (Figure~\ref{draine}) is a delta function.

To obtain the dust distribution at a given time, we consider all the dust particles created at previous times. Figure \ref{multi_prof} shows
the behavior of dust particles created at various time intervals for $\beta =4$, $M_{gas} = 50~M_\earth$, $H/r = 0.1$, and linear dust
creation.  The generations of particles were tracked to $T = 1.5$, and their respective surface densities were summed and azimuthally 
averaged at each time interval.  This simulation tracks the entire history of dust generation, but in practice, dust blowout eliminates 
from the ``current'' surface density profile all the contributions of dust created before the last half time unit (i.e., $T-0.5$).

\subsection{Simulation Results\label{model_results}}

We now present our models for the fly-by encounter with a composite disk of planetesimals, gas, and dust.  Representative simulations are shown 
in Figures~\ref{evolution}--\ref{results4}.  Although most of the models are shown at $T = 1.5$, the images and profiles are representative of 
$0.8 \lesssim T \lesssim 2.0$.  To give our distance and time scales physical meaning, we set the distance between HD~141569A and the barycenter of 
the companions at $T = 1.5$ equal to that observed in the ACS images (C03).  Thus, one distance unit corresponds to 718~AU and one time unit corresponds to 1800~yr. The companions are out of the frames.

For each of the Figures~\ref{evolution}--\ref{results4} we show the predicted logarithm of the density profile with linear dust generation (top row) or quadratic dust generation (bottom row). The use of the logarithmic stretch allow us to highlight faint features. The images are all normalized to the same arbitrary constant. In general, the color table of the bottom row has a lower limit value that the color table at the top row. This means that ``white'' in the bottom panels corresponds to somewhat less material than it does in the top panels. In this way, we can show faint features in the bottom rows that would not be visible were we to use the same color table as that from the top rows. This modification has little effect on the color the dark features, which implies that all panels within a given figure can be compared to each other. As mentioned before, an artificial inner hole has been set for the linear (at $r=0.1$) and quadratic simulations (at $r=0.15$). In each panel, the plot inset compares the measured density profile with the model density profile. The latter is normalized to the value measured at 500~AU.

Figure~\ref{evolution} shows the evolution of the dust structures for $\beta = 4$, $M_{gas} = 50~M_\earth$, $H/r = 0.1$, an initial disk size of one unit distance, and both linear and 
quadratic dust generation. Figures~\ref{results1} and 
\ref{results2} show the effect of the amount of gas on $\beta=4$ and $\beta=10$ dust particles, respectively.  The sharp, concentric rings in the gasless models 
(panels 1 and 4) are artifacts of discrete dust generation.  Figure~\ref{results3} illustrates the role of the gas temperature, and 
Figure~\ref{results4} shows the roles of disk sizes and planets.  A side-by-side comparison of the observed disk and our favorite model is presented 
in Figure~\ref{comparison}.  

All the models look similar.  Despite the dynamical importance of RP, the overall appearance of the dust disk resembles the underlying distribution of 
planetesimals. Figures~\ref{results1} and \ref{results2} show that models with gas reproduce the observed profiles better than models without gas, but it is not clear 
that $10~M_\earth$ of $H/r=0.1$ gas (panel 2 of each figure) is better than $50~M_\earth$ of $H/r=0.1$ gas (panel 3 of each figure).  There are no 
significant qualitative differences between the $\beta=4$ and $\beta=10$ conditions shown in Figures \ref{results1} and \ref{results2}, respectively, as 
expected from the results in \S\ref{conf}.  There is also no clearly preferred prescription for dust creation.  The quadratic prescription does produce a 
more disorganized structure, as the dust is mostly generated in the denser spiral arms. Overdensities in the planetesimal distribution produce streams of matter in these models.   Nevertheless, the combination of large $\beta$ and quadratic 
dust generation with $H/r=0.1$ gas (Figure~\ref{results2}, panels 5 and 6) reproduces very well the decay of the surface-density profile. Notice that because more gas confines the dust better, the spirals in panels 3 and 6 of each figure look darker than the others.

Figure \ref{results3} shows that the simulations are sensitive to the gas temperature.  Cold gas (panels 1 and 4) tends to produce sharper spiral 
structures than hot gas, because cold gas is denser and imparts greater drag on the dust.  This effect is less pronounced for quadratic 
dust generation, as the large density variations obscure the temperature-dependent effects of the gas.

All the models produce a two-armed spiral.  The arms are less pronounced for quadratic dust generation than for the linear dust generation because
the former emphasizes regions of large planetesimal density.  The density and extent of the arm diametrically opposite the companions can be reduced 
(and made more compatible with the observations) by diminishing the size of the initial disk (Figure~\ref{results4}).  The appearance of this arm can 
also be changed with a different encounter geometry. \citet{lar01} show that, in a encounter between a circumstellar disk and a passing star, the 
spiral arm opposite the projectile is affected by the indirect tidal component of the potential, which diminishes as the collision becomes more 
hyperbolic. 

In general, the models with quadratic dust generation produce more disorganized images than models with linear dust generation.  Planetesimal 
condensates are strong sources of dust that produce ``noise'' in the models.  However, they also produce more peaked surface-density profiles, which 
match the observed profiles better.  Flatter initial density profiles (not shown here) produce large amounts of dust at large distances.  Disks with initial radii larger than $\sim$1000~AU produce remnant structures that are not observed, including 
very long spiral arms.  Conversely, disks with an initial radius of 400~AU yield dust distribution that reproduce well the observed large-scale
arms and surface-density profiles (Figure~\ref{results4}).  We surmise from our models that HD~141569A's initial planetesimal and gas disks had 
radii $\lesssim 700$~AU and 10--50~$M_\earth$ of $H/r=0.1$ gas (Figure~\ref{comparison}).




\subsection{The Effect of Giant Planets}

From Figure~\ref{multi_prof} it seems clear that different dust configurations (i.e. large holes) can be created by manipulating the dust generation history of the system. The size of the hole can be controlled by varying $\beta$ or the time between dust generations.  Though theoretically possible, such contrived scenarios are physically unreasonable.

A more reasonable agent for creating a hole in the dust distribution is a planet (or planets) within $\sim150$~AU.  If large enough, planets would clear out
the planetesimals and the gas, and create a region without dust.   Each planet would sweep up a lane along its orbit as wide as a few times its Hill 
radius, $R_H$.  From \citet{bry00}, a $5~M_{Jup}$ planet with a semimajor axis of 100~AU and eccentricity $e=0.6$ would clear a region with a half-width of 90~AU. Alternatively, multiple planets in less eccentric orbits would have similar effect.  

Here we present exploratory simulations that include the effect of a giant planet. Panels~2 and 4 of Figure~\ref{results4} show models that include a $5~M_{Jup}$ planet in an eccentric orbit ($e\sim0.6$) around HD~141569A with a semi-major axis of 100~AU.  The planet clears the central region enough to greatly reduce the dust generation close to 
the star and produce an inner hole in the dust disk. Weak signatures of the eccentricity (the elongated shape and slight dust overdensities at the top and the bottom of the hole) are observed. The inner ring in the observed density profile is not reproduced in the observations, at least with these orbital parameters. Considering the uncertainties in the exact dust configuration close to the coronagraphic mask, this is a satisfactory result. A $5~M_{Jup}$ planet of age 5~Myr would be $\sim10^{-5}$ times less luminous, and over 12 magnitudes fainter in $K$, than HD~141569A \citep{bur97,gol04}.  Such an planet would be below the non-coronagraphic NICMOS detection limits \citep{kri98}.

The formation of small planets may produce sharp rings in the disk by stirring nearby planetesimals and enhancing dust generation. This process has been 
modeled in detail by \citet{ken04}. Both this effect and the presence of a giant planet modeled above produce inner ring variability on timescales of $\sim 1000$~yr.


\section{Discussion\label{model_disc}}

Our favored model of the scattered-light disk is shown in Figure \ref{comparison}.  This model assumes an initial disk size of 400~AU, 
$\beta =4$, and gas with with mass $50~M_\earth$ and an opening angle of $H/r=0.1$.  The model reproduces the large-scale spiral arms and, 
despite a large value of $\beta$, produces localized structures.  It does not reproduce the sharp and coherent rings or disk asymmetries of the disk around HD~141569A, as observed by C03, which suggests that a different encounter geometry should be considered.  The model includes a $5~M_{Jup}$ 
planet in an eccentric orbit ($e=0.6$) with apastron 160~AU for the purpose of clearing out a region of planetesimals close to the central 
star.  The characteristics of this planet are not tightly constrained by the model, and other configurations (including multiple planets) 
are possible.  The simulations clearly indicate that a mechanism capable of clearing the inside of the disk (like a planet or planets) is 
necessary to explain the observed density profile.  There may be additional observational evidence supporting the existence of a planet.  
\citet{bri02} have detected H$^+_3$ within 7~AU of HD~141569A, which they suggest may come from the extended atmosphere from a giant 
protoplanet.

Although there is favorable evidence for the existence of planet(s), we should also consider the effect of ice condensation on the surface 
density of the dust.  An abrupt change in the surface density is expected upon crossing the ``snow line,'' 
which for HD~141569A's disk is $\sim150$~AU (LL03).  The location of the snow line depends strongly on grain size, so it likely varies 
with observed wavelength.  Current coronagraphic images lack the spectral resolution needed to probe these variations.  While crossing 
the snow line cannot explain the paucity of dust close to the star, it likely affects the shape of the observed surface density. 

Most of the gas simulations do a similar job reproducing the optical depth images and profiles. It would seem that the only hard constraint that they provide is that the presence of gas is necessary. The quadratic dust generation method does a somewhat better job than the linear one in matching the density profiles. Interestingly, Figure~\ref{conclusion} shows that the planetesimal distribution itself can also match the outer surface density profile. The surface density of planetesimals starts as $r^{-1.5}$, but the truncation by the encounter makes it steeper by the time the simulations are examined. The surface density of dust generated quadratically also matches the distribution of planetesimals: while the dust is generated as the square of the planetesimal distribution, it is blown outward by radiation pressure. This process tends to flatten the simulated profile.
 
The applicability of our models to the debris disk around HD~141569A depends on the premise that the observed dust grains are small enough 
so that $\beta>1$. This condition is borne out by the disk colors, the mean-scattering phase, and thermal models.  Could larger particles 
($\beta<0.5$) be contributing to the observed scattered light image?  In deriving the scattering-opacity function (Figure~\ref{draine}),
we assumed that the collisional equilibrium timescale is smaller than any other systemic timescale, which implies that the number of 
grains of size $s$ decreases as $s^{-3.5}$.  If small grains were more quickly removed from their points of creation, then an excess of 
larger grains would result because they would no longer suffer erosion from collisions with the smaller grains \citep{the03}.  In the 
absence of gas, $\beta<0.5$ grains are eliminated by PR drag on timescales orders of magnitude larger than the RP timescales \citep{che01},
so the proportion of larger grains would further increase.  However, in the presence of modest amounts of gas, $\beta<0.5$ grains migrate 
to stationary orbits (TA01), and small particles settle down at the outer edge of the disk.  The timescale over which this migration
occurs is quite small for small grains, but larger than the purely dynamical timescales associated with $\beta>0.5$ motions.  All these 
processes may alter the shape of the dust size distribution.  If, for example, the distribution is flatter than $s^{-3.5}$, then the peak
of the scattering-opacity function is broadened.  While the optical image of the disk may still be dominated by $\beta$-meteoroids, the 
contributions of larger, bound grains may be significant.   

The entangled roles of RP, gas drag, and the dynamical role of the companions greatly complicate any effort to model all components of
the disk.  To do so successfully requires multi-wavelength images of the disk.  Observations by the {\it Spitzer Space Telescope} should
soon help this situation, as {\it Spitzer} will observe larger grains that are less affected by RP.  Future observations with the 
Atacama Large Millimeter Array (ALMA) will also contribute valuable information about the composition and distribution of the dust.

\section{Conclusions\label{conc}}

We have developed dynamical models to investigate the structures observed in the debris disk around HD~141569A.  The models include, for 
the first time, the effects of radiation pressure and gas drag.  The disk's colors and scattering phase indicate that the scattered-light 
images are produced by submicron-sized dust grains.  To understand these images, we must consider the behavior of grains with values of 
$\beta$ (the ratio of radiation and gravitational forces) larger than one.  These grains are expelled from the disk in shorter-than-dynamical
timescales.  We show that up to $60~M_\earth$ of gas may be present in the disk.  Gas drag slows the blow-out of grains and is fundamental 
to understanding the structure of the disk.

\begin{figure}
\plotone{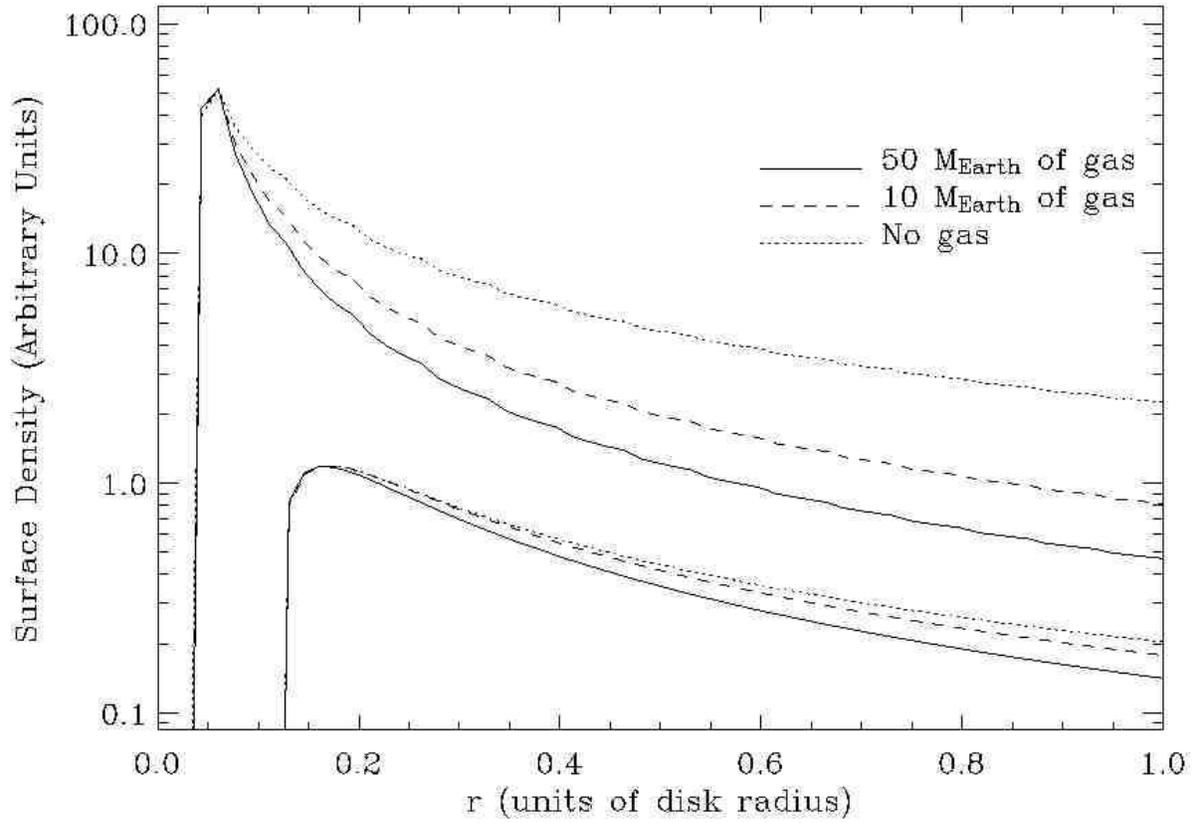}
\caption{\label{1d_model2}Steady state dust surface density as a function of distance for an isolated disk but with the number of dust particles generated proportionally to the square of the planetesimal surface density. Line traces as in Figure \ref{1d_model}. Notice that a broad ring is produced in all cases.}
\end{figure}

\begin{figure}
\plotone{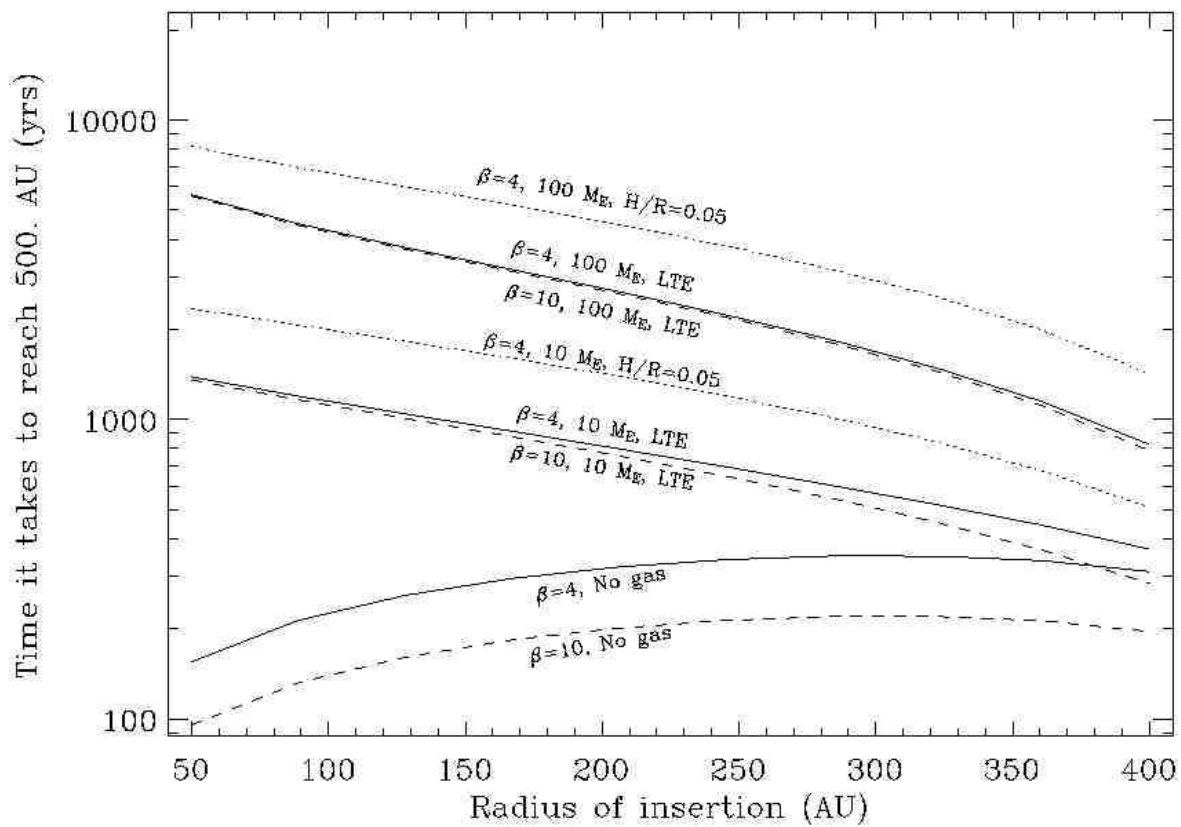}
\caption{\label{timescale}The time in years it takes for a dust particle inserted with keplerian velocity at different radii to reach 500 AU. The central star is 2.3 M$_{\sun}$. The gas sound speed is parametrized by the opening angle, H/r, where H is the disk thickness. Solid: $\beta=4$, gas in LTE; dashed: $\beta=4$, H/r=0.05; dotted: $\beta=10$, gas in LTE. For each set of plots, from bottom to top: no gas, 10 M$_\earth$, 100 M$_\earth$ (this is the gas amount within 500 AU).}
\end{figure}

\begin{figure}
\epsscale{.80}
\plotone{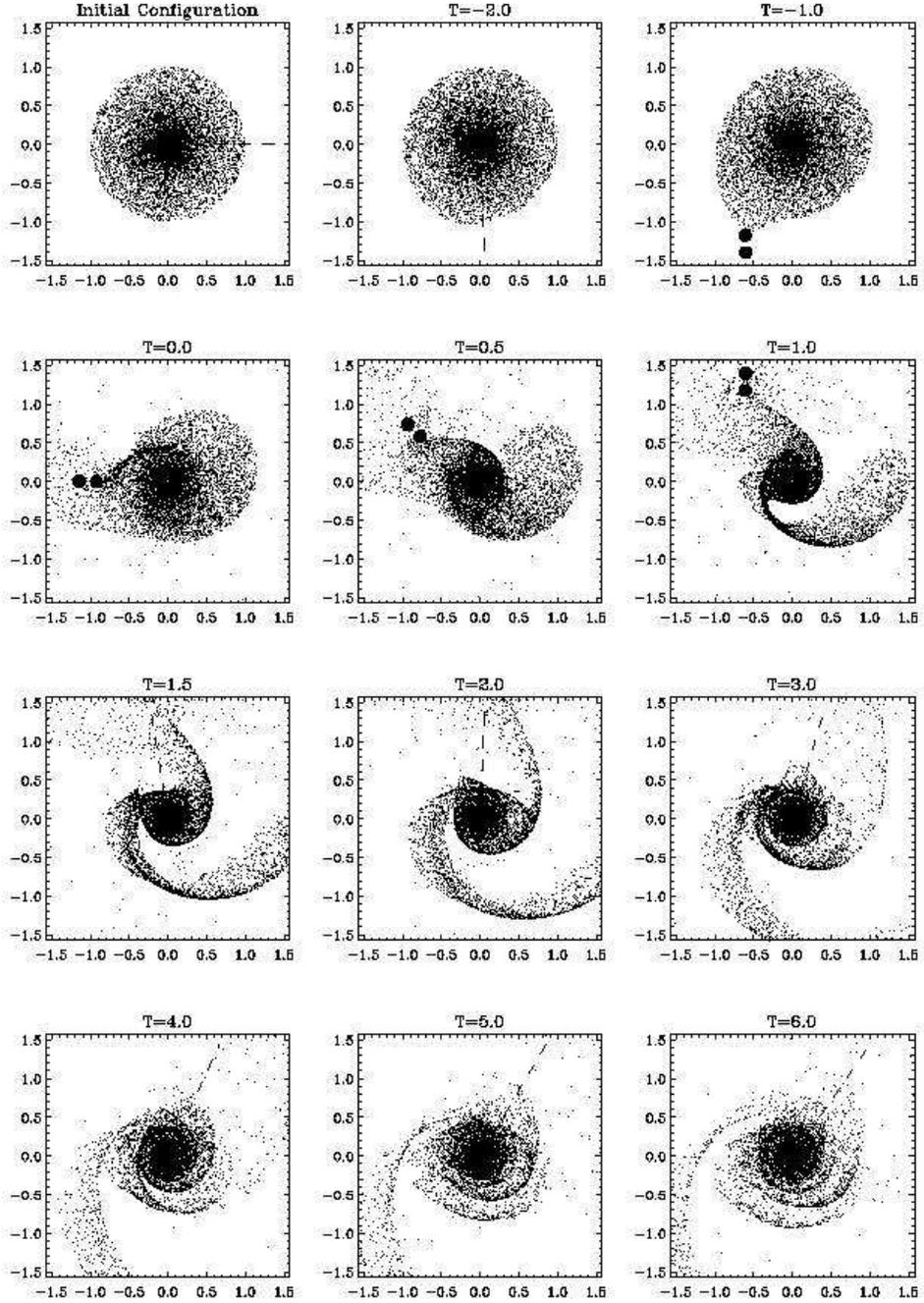}
\caption{\label{panel}Planetesimal evolution as a function of time, for a prograde encounter. The periastron occurs a one unit distance and in this particular example that is also the radius of the disk. The units of time are such that the period of a planetesimal at radius one is approximately 2$\pi$. The closest approach occurs at T=0.0, with the largest of the companions closest to the center. The initial surface density distribution goes as $r^{-1.5}$. The position of the binary is indicated with two filled circles or a dashed line (when outside the presented frame).}
\end{figure}

\begin{figure}
\plotone{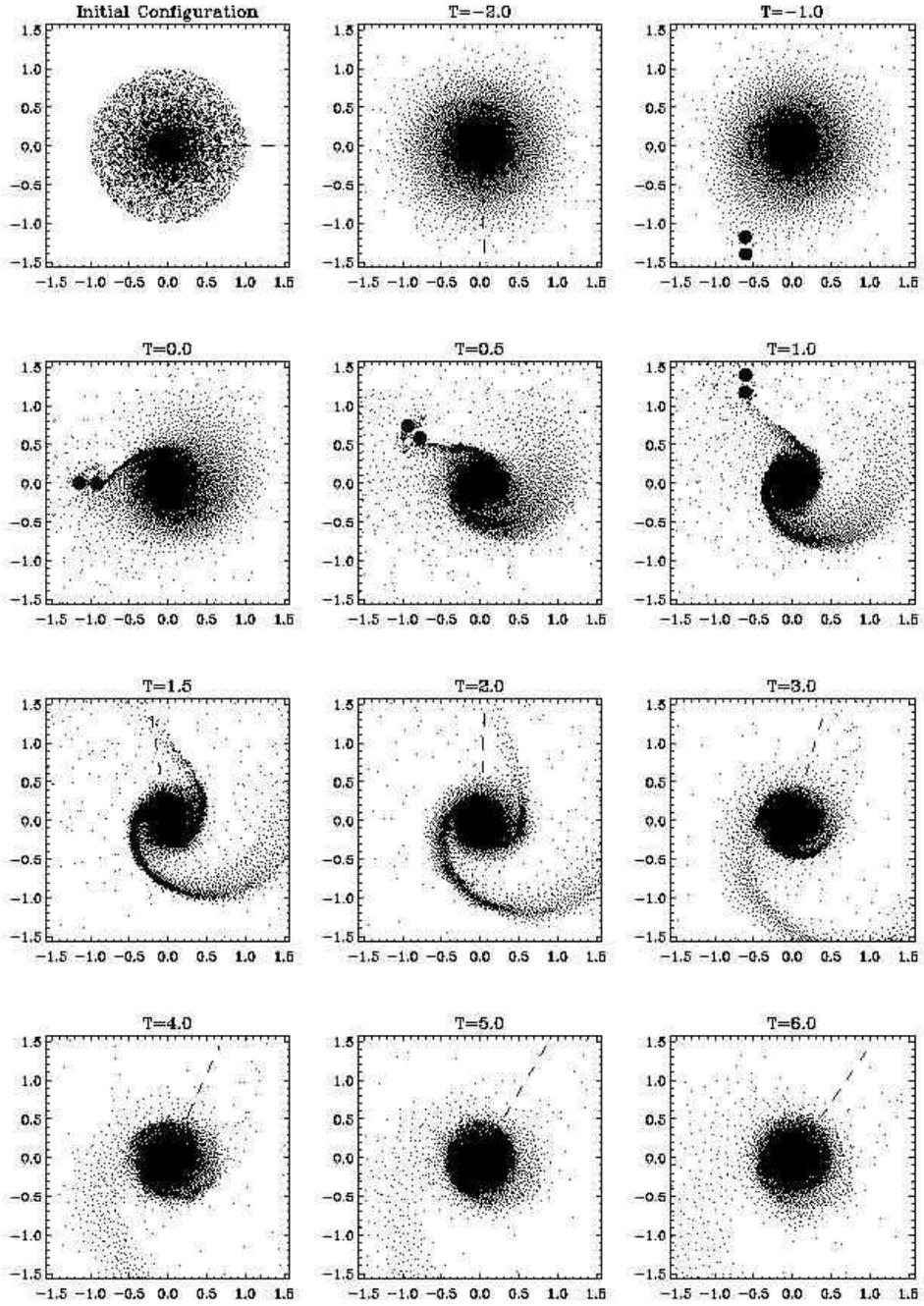}
\caption{\label{panel_gas}Gas evolution as a function of time. The gas has H/r=0.1. The parameters of the simulation are the same as that of Figure \ref{panel}.}
\end{figure}

\begin{figure}
\plottwo{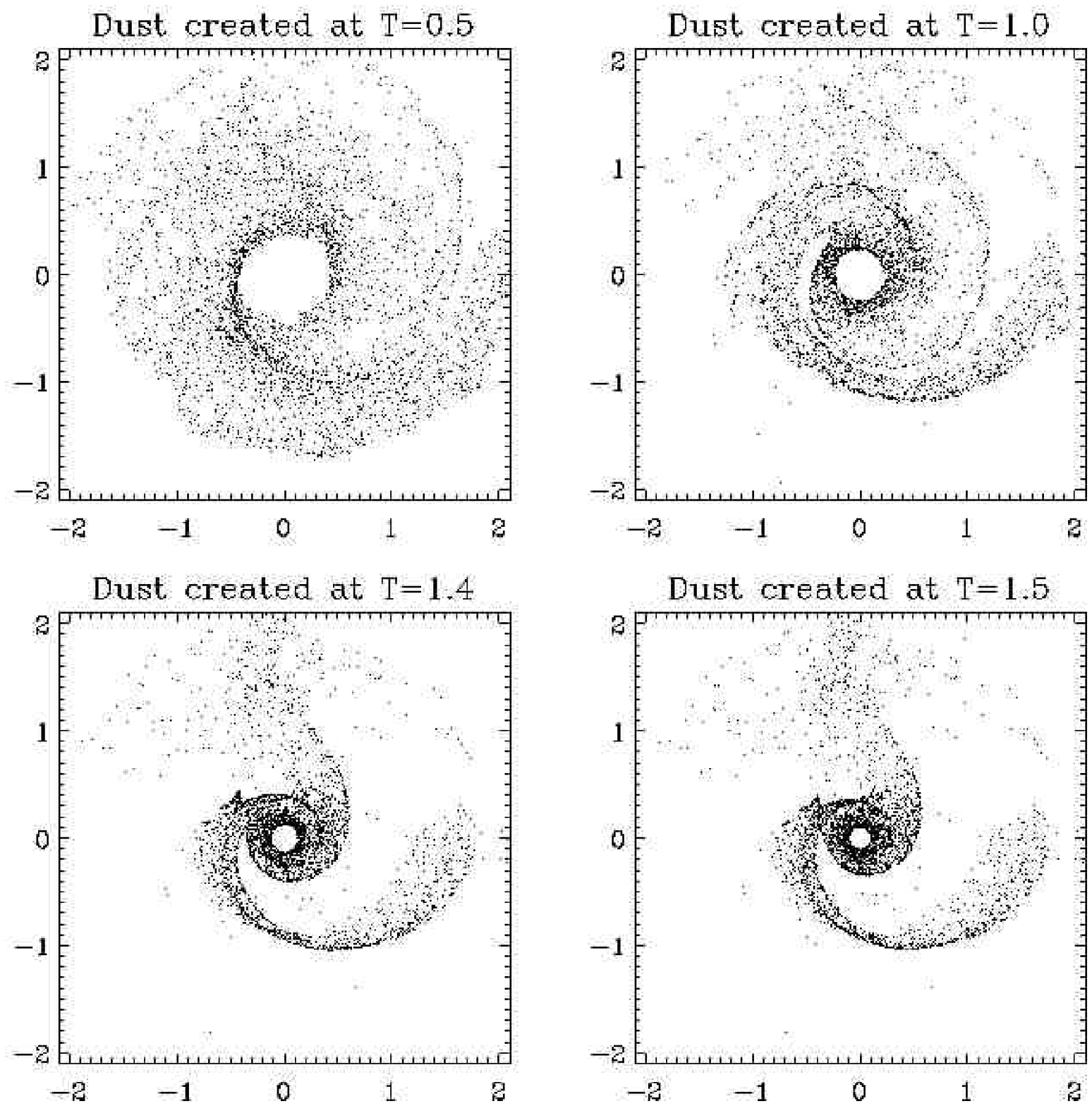}{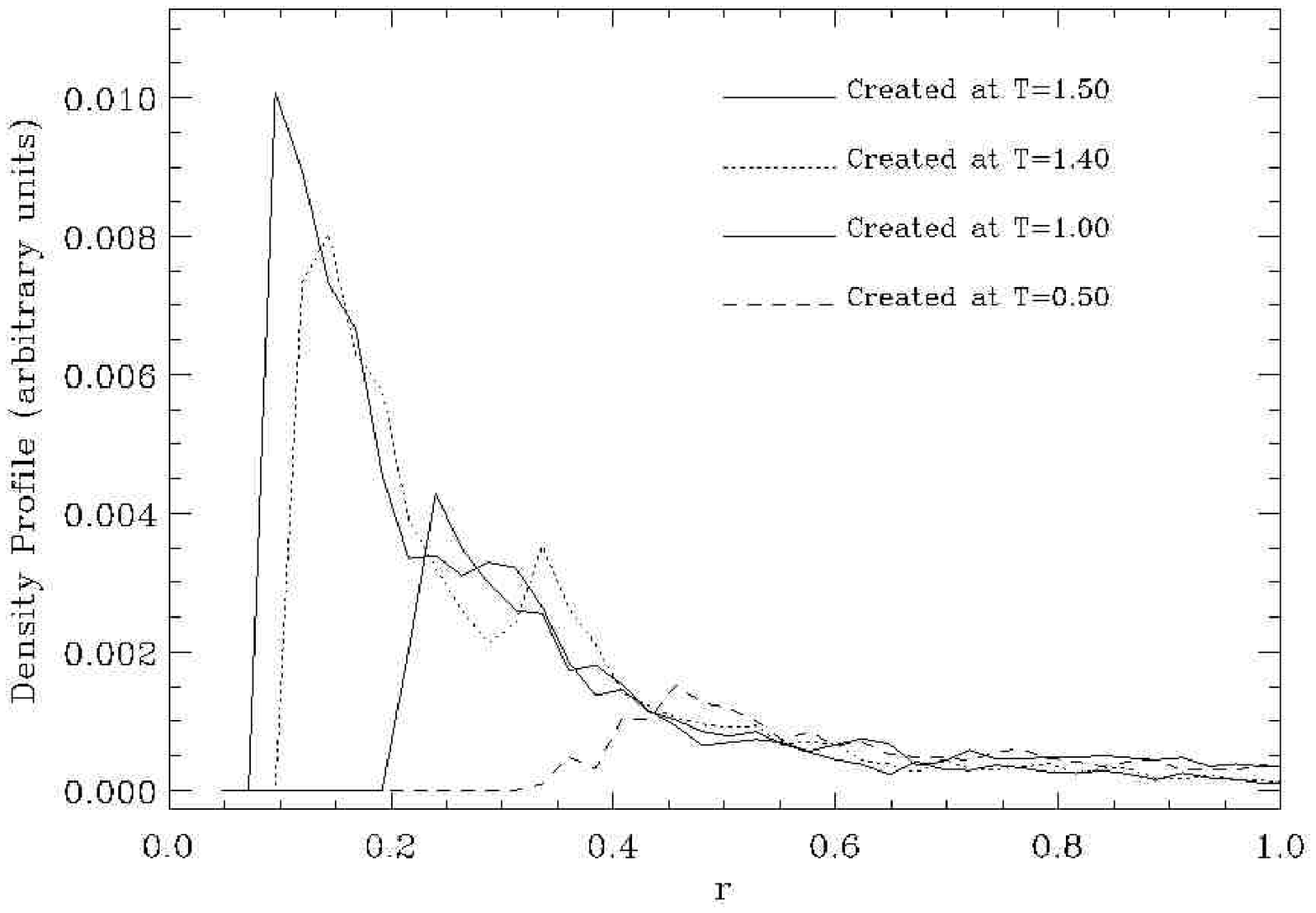}
\caption{\label{multi_prof}The observed profile at a given time is built by adding contributions from dust at previous times. In this example, $\beta=4$, M$_g$=50M$_\earth$, with H/r=0.1, $T=1.50$, quadratic dust generation. The panels show the configuration of dust created at previous times and observed at $T=1.50$. The plot shows the azimuthally-averaged densities. These plots show that the disk is mainly cleared from the inside out.}
\end{figure}

\begin{figure}
\plotone{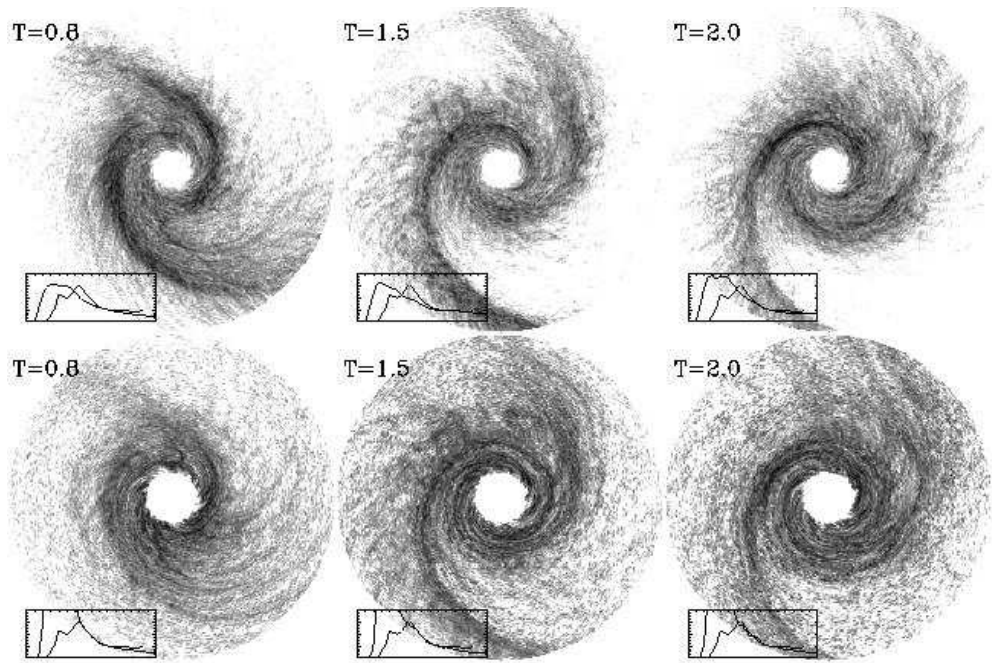}
\caption{\label{evolution}Time evolution of the dusty disk. The image shows the predicted density profile, azimuthally smoothed and gaussian convolved to match the measurements. The top row shows models with linear dust generation and the bottom row shows models with quadratic dust generation. The profiles have been normalized to the same arbitrary maximum value. The panels display the logarithm of the density profiles. To highlight faint structure, white color in the bottom panels corresponds to pixels 16 times fainter than in the top panels. The companions are on top, just outside the frame. The plot inset compares the measured density profile (thin line) with the model density profile (thick line). To set the distance scale, the periastron is set at 718 AU. In these simulations, that is also the size of the initial disk. All models have $\beta=4$, 50  M$_\earth$ of gas, $H/r=0.1$. Three times are shown: T=0.8, T=1.5, and T=2.0.}
\end{figure}

\begin{figure}
\plotone{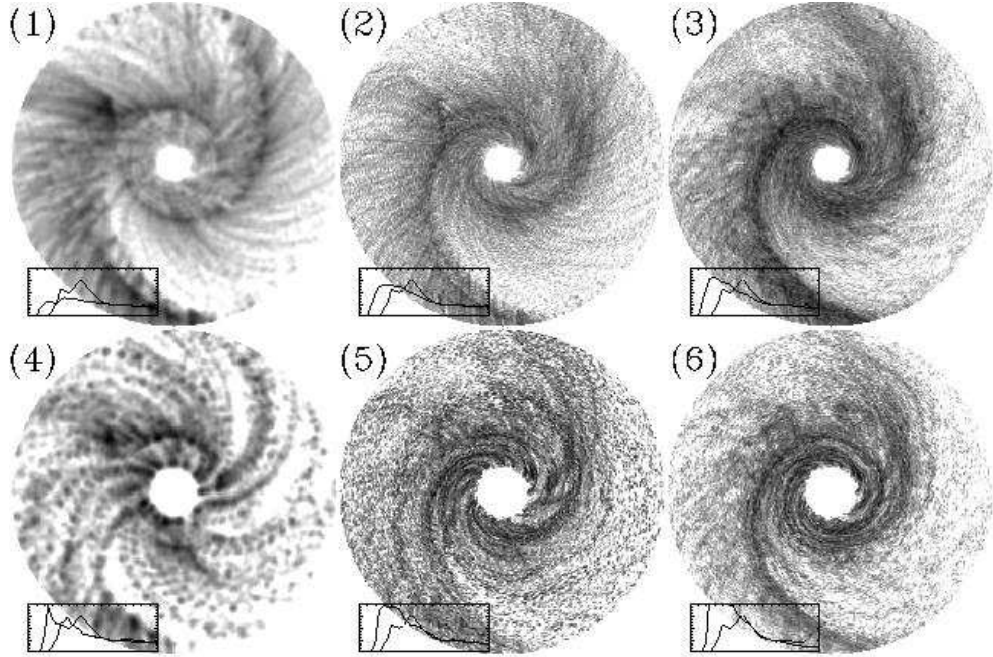}
\caption{\label{results1}The effect of the amount of gas on the dust distribution. Profiles have been processed as in Figure \ref{evolution}. Results of different model runs, for T=1.5. All models have $\beta=4$. The top row shows models with linear dust generation and the bottom row shows models with quadratic dust generation. White color in the bottom panels corresponds to pixels 4 times fainter than in the top panels. Top row, from left: (1) No gas, linear dust generation; (2) 10 M$_\earth$ of H/r=0.1 gas, linear dust generation; (3) 50 M$_\earth$ of H/r=0.1 gas, linear dust generation. Bottom row, from left: (4) No gas, quadratic dust generation; (5) 10 M$_\earth$ of H/r=0.1 gas, quadratic dust generation; (6) 50 M$_\earth$ of H/r=0.1 gas, quadratic dust generation. In order to correct for artifacts of the dust generation process, the gasless models (panels 1 and 4) have been further processed by taking 3-point medians in every pixel.}
\end{figure}

\begin{figure}
\plotone{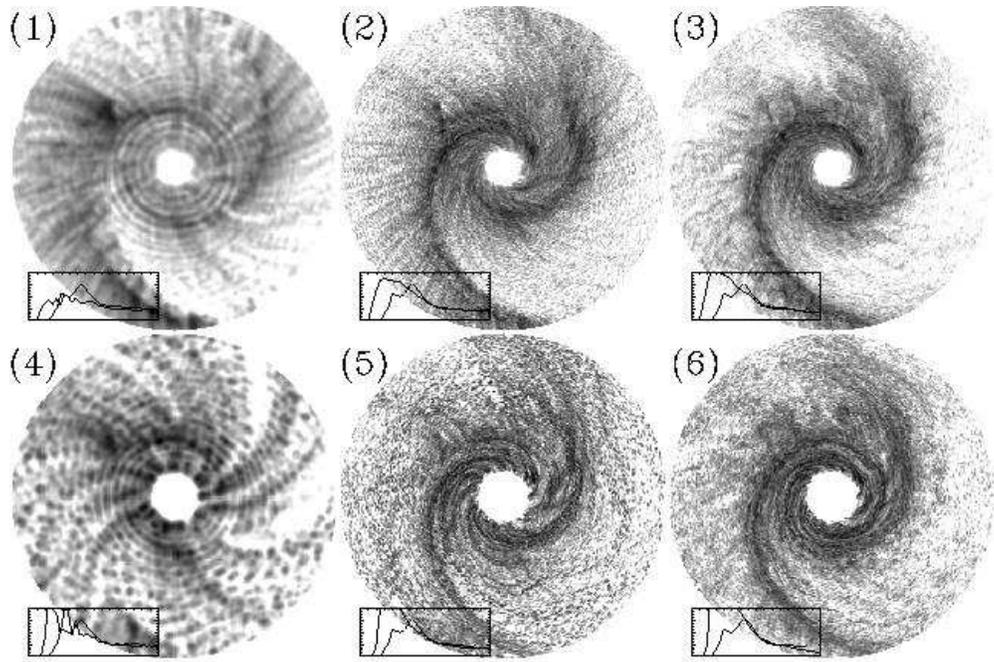}
\caption{\label{results2}The effect of the amount of gas on the dust distribution. Same as Figure \ref{results1} but with $\beta=10$. The concentric rings for the gasless models are artifacts from the dust generation process.}
\end{figure}

\begin{figure}
\plotone{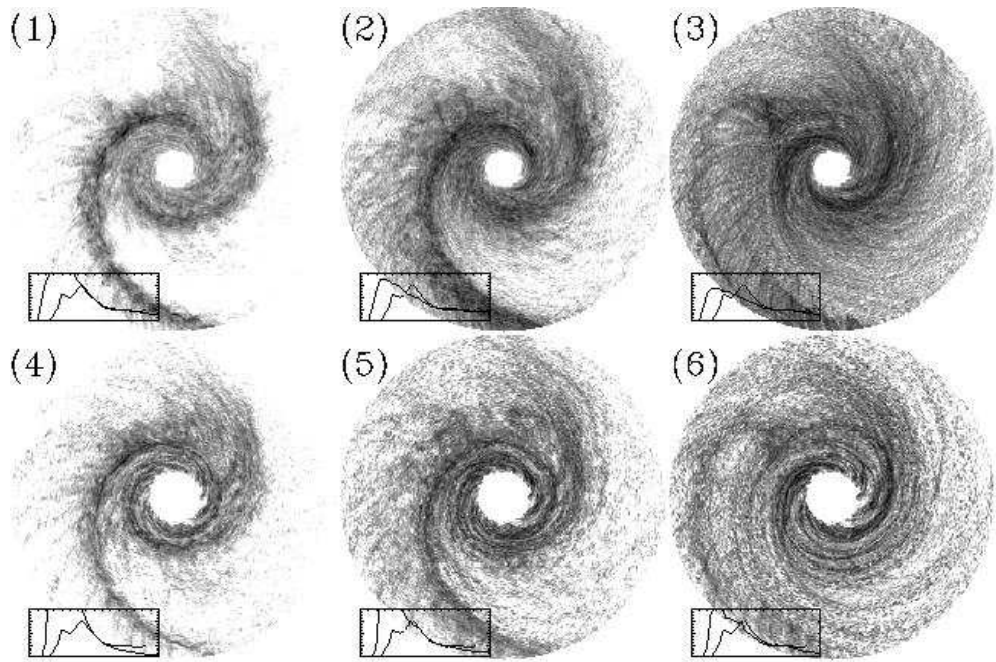}
\caption{\label{results3}The effect of the gas temperature on the dust distribution. All models have $\beta=4$ and 50 M$_\earth$ of gas. The top row shows the results of the linear dust generation method, and the bottom row the results of the quadratic one. Top row, from left: (1) H/r=0.05, linear dust generation; (2) H/r=0.1 gas, linear dust generation; (3) LTE, linear dust generation. Bottom row, from left: (4) H/r=0.05, quadratic dust generation; (5) H/r=0.1 gas, quadratic dust generation; (6) LTE, quadratic dust generation.}
\end{figure}

\begin{figure}
\plotone{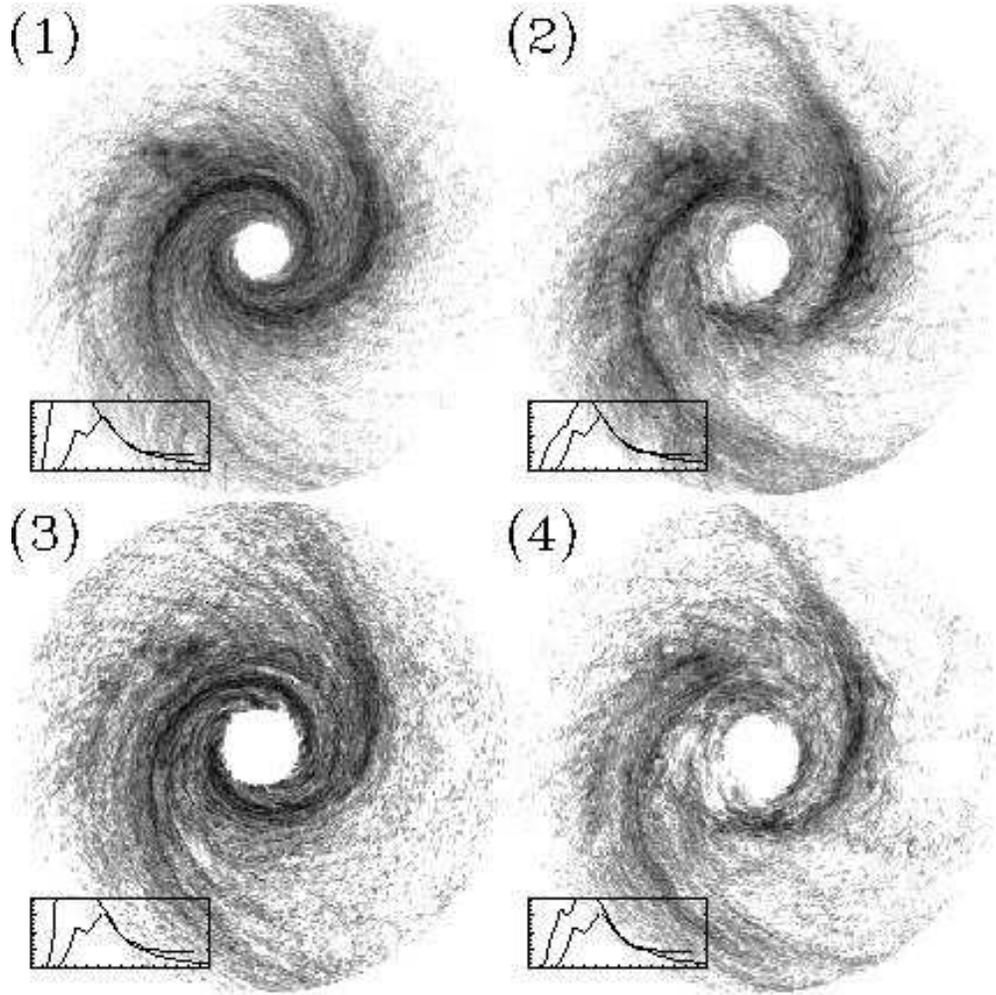}
\caption{\label{results4}Planets and small disks. The top row shows linear dust generation and the bottom row shows quadratic dust generation. All simulations have $\beta=4$, 50 M$_\earth$ of $H/r=0.1$ gas. (1) 400 AU initial disk, linear dust generation; (2)  400 AU initial disk, linear dust generation, with an eccentric ($e=0.6$), 5 M$_{Jup}$ planet with semimajor axis 100 AU; (3) 400 AU initial disk, quadratic dust generation; (4)  400 AU initial disk, quadratic dust generation, with an eccentric 5 M$_{Jup}$ planet within 160 AU; }
\end{figure}

If the dust is continuously created, its steady-state surface density is featureless and has a radial profile dependent upon the amount 
of gas and the prescription for dust creation.  To produce a broad ring of dust, its creation must be prevented within some inner radial
limit (Figure \ref{1d_model2}).  In principle, one can use the observed surface-density profile to estimate the amount of mass lost from
by the disk throughout its history.  However, because large grains remain in the disk longer than smaller grains, this estimation implies
an implausible amount of solid material in the disk.  Consequently, either the observed rate of dust generation is limited to times near 
the encounter of the disk with HD~141569A's binary companions, or small grains are created highly preferentially over larger ones.

Inspired by disk's large-scale spiral arms, we explored the effect on the disk from a parabolic fly-by of the binary companions.  Such an
encounter is consistent with the measured proper motion of the system.  Our models consider the dynamical evolution of three different 
populations: planetesimals (subject only to gravity), gas (subject to gravity and hydrodynamic forces), and dust (subject to gravity, 
radiation pressure and gas drag).  The Poynting-Robertson effect, self-gravity, and the dynamical effects of collisions are ignored. 
We vary $\beta$, the initial gas mass, the gas disk opening angle, and the prescription for dust generation.  We find that the dust 
distribution resembles that of the contemporary planetesimal distribution.  The models are most sensitive to the gas temperature, with 
hotter gas producing more disorganized structures than colder gas.  Five-fold differences in the gas mass and two-fold differences in 
$\beta$ do not produce significant differences in the simulated disks.  Quadratic dust generation produces more disorganized and less
distinct structures than linear dust generation.  Smaller initial disks produce smaller and less pronounced spiral arms. 
 
\begin{figure}
\plotone{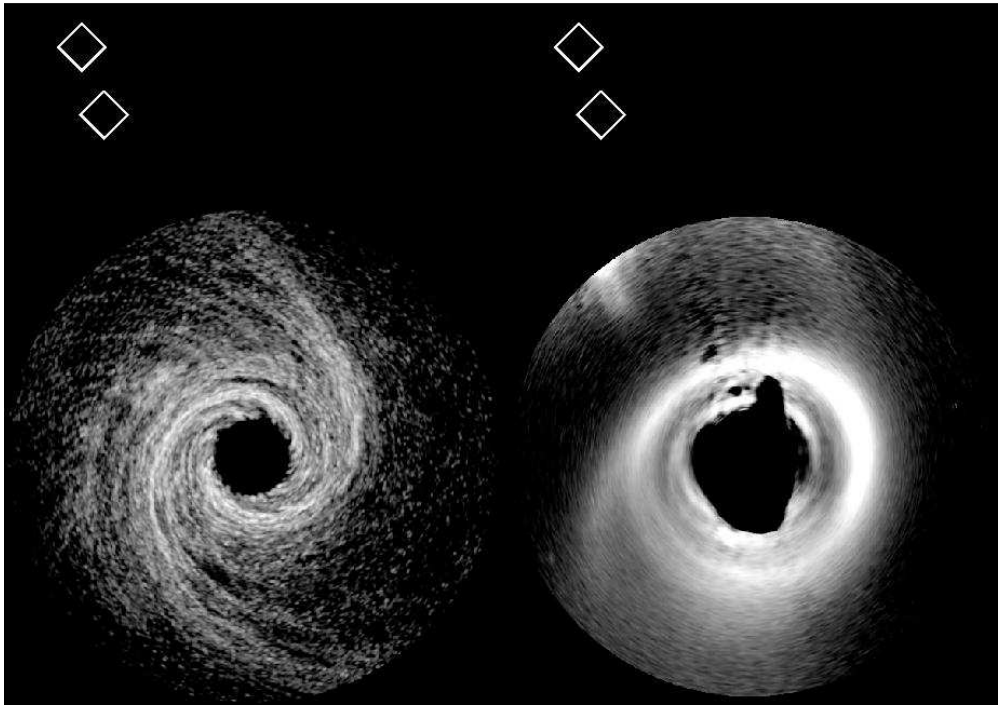}
\caption{\label{comparison}Comparison between best model and observations. Both are logarithmic stretches. The model ($\beta=4$, 50 M$_\earth$ of $H/r=0.1$ gas, cuadratic dust generation, with an eccentric 5 M$_{Jup}$ planet within 100 AU) has been rotated counterclockwise by 13 degrees.}
\end{figure}

\begin{figure}
\plotone{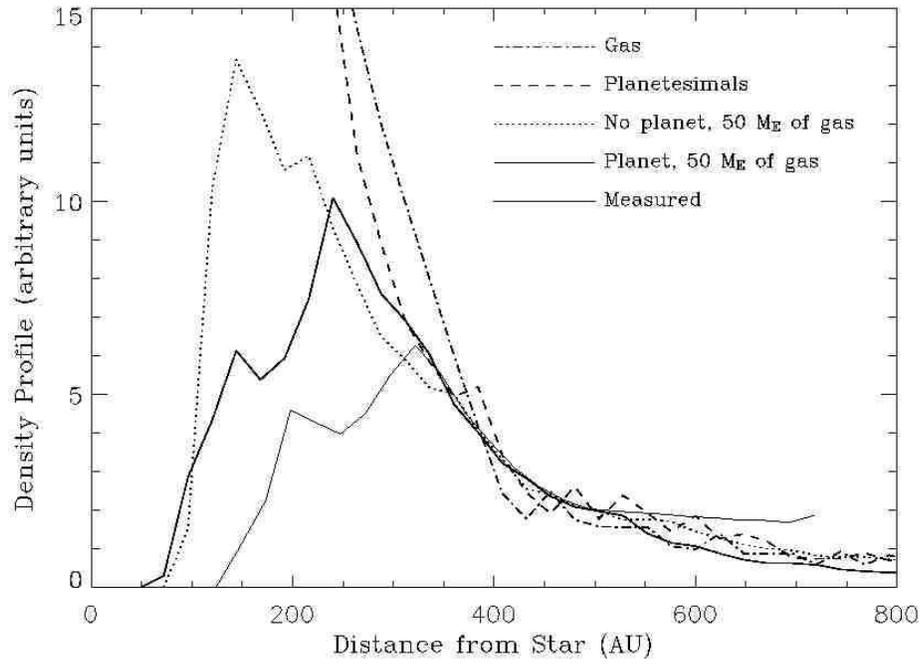}
\caption{\label{conclusion}Comparison between measured and modeled profiles, at $T=1.50$. The traces show density profiles generated quadratically in small disks ($\beta=4$, 50 M$_\earth$ of $H/r=0.1$ gas), with and without planets (Figure~\ref{results4}, panels 3 and 4). Also shown are the planetesimal and gas surface densities. These are steeper than initially set because of the truncating effect of the encounter.}
\end{figure}

Our favored model (Figure~\ref{comparison}) has an initial disk size of 400~AU, $\beta =4$, and gas with with mass $50~M_\earth$ and an 
opening angle of $H/r=0.1$.  A planet is introduced to reduce the surface density of the planetesimals close to the central star, but 
the characteristics of this planet are not tightly constrained.  The model successfully reproduces the large-scale structures seen in
the scattered-light images of the disk.  However, the model does not reproduce the tightly-wound spiral rings or other observed asymmetries, 
which suggests that another population of grains (perhaps with smaller $\beta$) or a non-coplanar encounter geometry should be
investigated.



\acknowledgments
The authors thank B.~Zuckerman, T.~Forveille, and J.~Kastner for generously providing the data from their 1995 paper.
We are grateful to K.~Anderson, W.~J.~McCann, S.~Busching, A.~Framarini, and T.~Allen for their invaluable contributions to the ACS project at JHU. 
ACS was developed under NASA contract NAS 5-32865, and this research has been supported by NASA grant NAG5-7697. We are grateful for an equipment grant from Sun Microsystems, Inc.
The Space Telescope Science Institute is operated by AURA, Inc., under NASA contract NAS5-26555. Furthermore, we gratefully acknowledge support from NASA Origins of Solar Systems grants NAG5-10732 and NNG04GG50G.  Finally we wish to thank the anonymous referee whose comments greatly improved the paper.


\clearpage

\clearpage

\clearpage

\end{document}